\documentclass[11pt]{article}
\pdfoutput=1
\usepackage{hyperref}
\hypersetup{
    colorlinks=true, 
    linktoc=all,     
    linkcolor=blue,  
    citecolor=red,
    filecolor=green,
    urlcolor=blue}
\usepackage{amsmath}
\usepackage{amssymb}
\usepackage{amsfonts}
\usepackage{graphicx,bbm,mathrsfs}
\usepackage{nicefrac}
\usepackage{bbm}
\usepackage[top=2.5 cm, bottom=4 cm, left=3 cm, right=3 cm]{geometry}     
\usepackage{color}
\usepackage{verbatim}

\newcommand{\be}{\begin{equation}}  
\newcommand{\ee}{\end{equation}}  
\newcommand{\bea}{\begin{eqnarray}}  
\newcommand{\eea}{\end{eqnarray}}

\newcommand{\tr}{\operatorname{tr}}

\newcommand\lsim{\mathrel{\rlap{\lower4pt\hbox{\hskip1pt$\sim$}}
    \raise1pt\hbox{$<$}}}
\newcommand\gsim{\mathrel{\rlap{\lower4pt\hbox{\hskip1pt$\sim$}}
    \raise1pt\hbox{$>$}}}

\newcommand{\captionfonts}{\small}

\newcommand{\approptoinn}[2]{\mathrel{\vcenter{
  \offinterlineskip\halign{\hfil$##$\cr
    #1\propto\cr\noalign{\kern2pt}#1\sim\cr\noalign{\kern-2pt}}}}}

\makeatletter  
\long\def\@makecaption#1#2{%
  \vskip\abovecaptionskip
  \sbox\@tempboxa{{\captionfonts #1: #2}}%
  \ifdim \wd\@tempboxa >\hsize
    {\captionfonts #1: #2\par}
  \else
    \hbox to\hsize{\hfil\box\@tempboxa\hfil}%
  \fi
  \vskip\belowcaptionskip}
\makeatother   

\begin{document}

\begin{flushright}LAL 17-024  \\ LPT-Orsay-17-04\end{flushright}  

\vspace*{1cm}

\begin{center}

\thispagestyle{empty}

{\Large\bf 
Scalar Production in Association with a Z Boson at LHC and ILC: 
the Mixed Higgs-Radion Case of Warped Models 
}\\[10mm]

\renewcommand{\thefootnote}{\fnsymbol{footnote}}

{
\large Andrei~Angelescu$^{\,a}$~\footnote{andrei.angelescu@th.u-psud.fr}, 
\large Gr\'{e}gory~Moreau$^{\,a}$~\footnote{gregory.moreau@th.u-psud.fr}, 
\large Fran\c{c}ois~Richard$^{\,b}$~\footnote{richard@lal.in2p3.fr}
}\\[10mm]

\addtocounter{footnote}{-3}

{\it
$^{a}$~ Laboratoire de Physique Th\'{e}orique, B\^at. 210, CNRS,
Univ. Paris Sud, \\  Universit\'{e} Paris-Saclay, F-91405 Orsay Cedex, France \\
$^{b}$~ Laboratoire de l'Acc\'{e}l\'{e}rateur Lin\'{e}aire, IN2P3/CNRS,
Univ. Paris Sud, \\  Universit\'{e} Paris-Saclay, B.P. 34 F-91898 Orsay Cedex, France \\
}

\vspace*{12mm}

{  \bf  Abstract }
\end{center}

\noindent 
The radion scalar field might be the lightest new particle predicted by extra-dimensional extensions of the Standard Model. It could thus
lead to the first signatures of new physics at the LHC collider. We perform a complete study of the radion production in association with 
the Z gauge boson in the custodially protected warped model with a brane-localised Higgs boson 
addressing the gauge hierarchy problem. Radion-Higgs mixing effects are present. Such a radion production receives possibly resonant 
contributions from the Kaluza-Klein excitations of the Z boson as well as the extra neutral gauge boson (Z'). All the exchange and mixing
effects induced by those heavy bosons are taken into account in the radion coupling and rate calculations. The investigation of the considered 
radion production at LHC allows to be sensitive to some parts of the parameter space but only the ILC program at high luminosity would cover most 
of the theoretically allowed parameter space via the studied reaction. Complementary tests of the same theoretical parameters can be realised 
through the high accuracy measurements of the Higgs couplings at ILC. The generic sensitivity limits on the rates discussed for the LHC and ILC  
potential reach can be applied to the searches for other (light) exotic scalar bosons.

\clearpage

\tableofcontents

\newpage

\section{Introduction}  \label{se:intro}

After the discovery of the Higgs boson and the completion of the Standard Model (SM), the search for new particles at the Large Hadron Collider (LHC) is more and more intense. 
Precise measurements of Higgs couplings are the natural complement of these direct searches given that Higgs couplings could be influenced by virtual 
exchanges and/or mixing effects of exotic particles. Interestingly, new scalar fields ($S$), arising in various SM extensions, could both be directly produced 
and mix with the Higgs boson. Such scalars can still be as light as a few tens of GeV given that for example the vanishing sensitivity of the LEP collider searches when the $ZZS$ 
coupling (to the $Z$ boson) reaches $\sim 1/10$ of the $ZZh$ (Higgs) coupling. LHC searches for scalars also suffer from limited sensitivity to light scalars; for instance the 
powerful investigation performed in the diphoton decay channel becomes inefficient for masses below $\sim 60$~GeV given the trigger limitations.
The future $e^+e^-$ International Linear Collider (ILC) and CLIC, which shall collect more than 100 times the LEP luminosities and reach the TeV scale, 
are expected to improve the low scalar mass searches.

From the theoretical point of view, the warped extra dimension scenario proposed by L.Randall and R.Sundrum (RS)~\cite{RS} with 
a Higgs boson localised at (or close to) the TeV-brane, being dual to composite Higgs models~\cite{composite},
remains one of the most attractive extensions of the SM. In particular due to its elegant solution of the
the gauge hierarchy problem and its simple geometrical explanations of the fermion mass hierarchies~\cite{exRSflav1,exRSflav2}~-- in 
case of matter in the bulk. The RS paradigm -- including the dual composite Higgs scenarios -- constitutes an alternative, to the 
supersymmetric SM extensions, of a completely different nature. Nevertheless, both these kinds of SM extensions predict the existence of new scalar particles which 
could lead to clear experimental signatures at colliders. In the case of warped models, a predicted scalar is the so-called radion, which corresponds
to the dilaton field through the $AdS/CFT$ correspondence.

The phenomenology at colliders of the RS scenario is guided by the indirect constraints on the masses of the various Kaluza-Klein (KK) excitations.
Let us thus shortly review the constraints on such a scenario arising from the electroweak precision tests (EWPT).
In the RS model with a custodial symmetry gauged in the bulk~\cite{RScusto}, the bounds from EWPT can be reduced down
to gauge boson masses $m_{KK} \gtrsim 3-5$~TeV~\cite{BouchartEW,MalmEW} for the first KK excitation of say the photon, in case of a purely 
brane-localised Higgs~\footnote{$\ \sim 3$~TeV for a bulk Higgs localised towards the TeV-brane~\cite{SridharEW}.}. In RS versions with a bulk Higgs field 
unprotected by a custodial symmetry, these bounds become $m_{KK} \gtrsim 7.5$~TeV 
for a Higgs profile still addressing the gauge hierarchy problem ($\beta = 0$)~\cite{SridharEW,NeubertEW}~\footnote{$\ \sim 2$~TeV~\cite{SridharEW}
with a deformed metric, with deviations from AdS geometry near the Infra-Red (IR) brane~\cite{Quiros}.}, 
and, $m_{KK} \gtrsim 13.5$~TeV for the brane-Higgs limit ($\beta \to \infty$)~\cite{NeubertEW}. 
\\ 
In contrast, within custodially protected warped models, the lightest KK excitations of fermions (custodians) can reach masses as low as the TeV scale 
while satisfying the EWPT affected by their loop contributions to the oblique parameters S,T~\cite{KKfloop} or their direct (mixing) corrections to the $Zbb$ 
vertex~\cite{BouchartEW}.

The radion scalar field, corresponding to the fluctuations of the metric along the extra dimension, has a typical mass around the EW energy scale~\cite{RadionGW}, 
within the standard mechanism of radius stabilisation based on a bulk scalar field~\cite{GW}. The EWPT [via the S,T,U parameters] and LEP limits allow radion 
masses between $\sim 10$~GeV and the TeV scale, depending on the curvature-scalar Higgs mixing (for SM fields on the IR brane)~\cite{RadionEWPT}. 

Given those mass bounds, the radion might be the lightest new particle and thus appear as the first signature of warped models at colliders~-- 
before KK fermion~\cite{KKfermionLHC} or KK gauge boson~\cite{KKbosonLHC} productions. The detection of the radion would constitute the discovery of
a second scalar field, after the Higgs boson observation. This new boson should then be disentangled from other scalar particles predicted by supersymmetric
models or other scenarios with extended Higgs sectors. 

The radion is mainly produced at LHC by gluon-gluon fusion (see e.g. Ref.~\cite{Frank} for a recent paper) but some 
model-dependence might affect this process as we discuss now. The LHC data~\cite{ATLASweb,CMSweb}  
on the Higgs rates~\footnote{These data constrain the Higgs-radion mixing to be small enough to recover a SM like Higgs boson.} lead to 
$m_{KK} \gtrsim 11$~TeV for a brane-Higgs~\footnote{$\ \sim 7.25$~TeV for a narrow bulk-Higgs.} 
within a custodially protected RS model~\cite{NeubertLHCH}.~\footnote{Those limits hold for a maximal absolute value $y^*=1.5$ of the anarchic dimensionless 5D 
Yukawa coupling constants, and are even more severe for a larger value $y^*=3$.} These constraints arise essentially because of the contributions of KK modes to the 
Higgs production reaction with the highest cross section: the loop-induced gluon-gluon Fusion (ggF) mechanism (see e.g. Ref.~\cite{BggF}). 
To reduce this limit on the KK scale $m_{KK} $
down to the TeV scale (comparable with EWPT limits), and in turn reconcile the related gravity scale at the IR brane with the fine-tuning problem, one may expect
some new physics effects (brane-localised kinetic terms, different fermion representations under the custodial symmetry, cancellation mechanisms\dots) in the 
triangular loop of the ggF mechanism, suppressing the KK mode contributions. This introduces some unknown model-dependence in the Higgs ggF mechanism which 
would also affect the similar ggF process of the radion scalar production.  

In contrast, the Higgs ($h$) production in association with an EW gauge boson ($V\equiv Z,W$), 
followed by the Higgs decay into a pair of bottom quarks, induces -- due to KK 
mixing~\cite{BouchartZZH} -- a limit of $m_{KK} \gtrsim 2.25$~TeV ($3.25$~TeV) with $y^*=1.5$ ($y^*=3$) for a brane-Higgs [and
slightly above for a narrow bulk-Higgs] still in custodial warped models~\cite{NeubertLHCH}. Such values are acceptable from the fine-tuning point of view. 
Hence there is no strong reason to assume that this tree-level $hV$ production is sensitive to unknown effects.
A similar conclusion then holds for the Radion ($\phi$) production in association with a gauge boson $V$.
 
The $\phi Z$ production in particular 
possesses other interests in some regions of the RS parameter space. For example, the radion discovery at LHC through its ggF production
is challenging if the radion mass satisfies $m_{\phi}<2m_Z$, closing kinematically the golden channel $\phi\to ZZ$~\footnote{Below this threshold, the channel 
$\phi\to ZZ^*$, into a virtual Z boson, may still allow to reconstruct one on-shell Z boson decaying to charged lepton pairs.}, and is too small to allow for the detection 
of the diphoton decay $\phi\to \gamma\gamma$. 
The $\phi Z$ production would then offer an additional on-shell Z boson which helps for the tagging of the final state. 
Another situation motivating the $\phi Z$ production search is a suppression of the ggF rate due to a significant decrease of the radion coupling to gluons
as occurs in some parameter regions.~\footnote{$m_\phi\gtrsim 200$~GeV and $\xi = {\cal O}(1)$, as 
shown in Ref.~\cite{Frank} (where the effect of the coloured KK fermions on the $\phi gg$ loop is neglected).}

Regarding the future $e^+e^-$ ILC machine, 
the $\phi Z$ production would be the dominant radion production mode~\cite{FRnote}, similarly to the Higgs boson case.
The $\phi Z$ production in a leptonic machine 
is also an important channel because, as for the $hZ$ channel, it allows for a decay independent search -- based on the simple $2\to 2$ body kinematics --
that should permit in particular to cover low radion masses being challenging at LHC.

Therefore, in this paper, we study the $\phi Z$ production in custodially protected warped models with a brane-localised Higgs boson. The analytical 
calculations of the radion couplings allow us to compute the complete $\phi Z$ production cross section, both at the LHC and ILC colliders. 
The LHC and ILC turn out to constitute complementary machines in regard to the $\phi Z$ investigation.
The $\phi Z$ reaction proceeds through the s-channel exchange of the EW $Z$ boson, its KK excitations as well as the extra $Z^{\prime}$ gauge boson (issued 
from the extended bulk custodial symmetry). All these contributions together with their interferences are taken into account.
The effects of the various KK mixings in the radion couplings and KK exchanges in the s-channel are 
discussed, as well as the possibility to reconstruct the invariant mass of the first two resonant heavy boson eigenstates (mainly KK modes)  
almost degenerate in mass. Such a spectacular resonance observation would constitute a double discovery of the radion and first KK gauge bosons. 
The resonant KK gauge boson detection through its decay to $hZ$ is also quantitatively studied. Indeed, 
the $\phi Z$ and $hZ$ productions should be consistently analysed together due to the $\phi-h$ mixing.
In view of the obtained $\phi Z$ and $hZ$ rates, we discuss the possibilities of experimental observations which rely on favoured radion decays, 
depending on the parameter space and in particular on $m_{\phi}$ values. 

Furthermore, we propose in the present work a more general experimental technique to search for an inclusive
final state $Z+X$ [where $X$ represents any SM or new particles], followed by the decay $Z \to 2$~charged leptons, based 
on a cut on the distribution of the $Z$ boson transverse momentum. The choice of the decay $Z \to \mu^+ \mu^-$ being a tagging device to allow trigger and detection.
Such a technique could also be applied for $ X\equiv \phi$ in RS versions 
different from the present one, e.g. with lower resonant KK $Z$ masses and/or favoured gluon decays for the radion 
(so that the associated tagged $Z$ becomes crucial for the detection). See for instance Ref.~\cite{FlavUnivRS} for a recent warped model of this kind.

At this stage we also mention the related work on the search of the radion at colliders~\cite{Frank} as well as the more general literature on the radion
phenomenology in warped scenarios with SM fields at the TeV-brane~\cite{RadSMatIR}, with only the Higgs boson stuck on the IR brane~\cite{RadHatIR} 
or the whole SM field content propagating in the bulk~\cite{RadSMinB}. Besides, there exists a connected study on the $hZ$ production through resonant 
neutral KK gauge bosons~\cite{AgaGopa}.

The paper is organised as follows. In Section~\ref{sec:couplings}, we present all the radion and Higgs couplings and calculate the KK mixing effects 
-- applying the so-called mixed KK decomposition to the gauge boson sector. Then we provide the analytical and numerical results for the $\phi Z$ and $hZ$ (Section~\ref{sec:RZ}) production 
cross sections at the LHC and ILC. The behaviours of these rates along the theoretical parameter space are explained there. 
In Section~\ref{sec:EXP}, experimental methods are proposed to detect the radion and/or (extra) KK gauge bosons. We conclude in the last section.

\section{Radion and Higgs Couplings} \label{sec:couplings}

\subsection{Model Description}

Our model is the RS scenario with the Higgs doublet localised on the IR brane, while the remaining fermionic and gauge fields are propagating in the bulk. 
The SM fermion mass hierarchy is generated through their wave function overlaps with the Higgs boson, as usually in this framework. 

In the $(+----)$ convention that will be used throughout this work, the well-known RS metric reads
\begin{equation}
{\rm d} s^2 = {\rm e}^{-2 k\,y} \eta_{\mu\nu} {\rm d}x^{\mu}{\rm d}x^{\nu} - {\rm d}y^2 \equiv g_{MN} {\rm d}x^{M}{\rm d}x^{N},
\label{static_metric}
\end{equation}
where upper case roman letters refer to 5D Lorentz indices and greek letters to 4D indices and $k$ being the 5D curvature scale, which is typically of the order of the Planck scale. The $y$ coordinate, which parametrizes the position along the extra-dimension, spans in the interval $[0,L]$. Throughout this work, we will consider that $kL$, the so-called volume factor, is equal to 35, such that the hierarchy problem is addressed. For the time being, we denote by $g_{MN}$ the \emph{unperturbed} metric, and postpone the inclusion of the scalar fluctuations for subsection~\ref{sec:HRcoupl}.

We consider the custodial gauge symmetry implementation with a Left-Right Parity~\cite{O3} as well as a more general implementation allowing potentially to address the $A_{\rm FB}^b$~\cite{RSAFBb} and $A_{\rm FB}^t$~\cite{RSAFBt} anomalies. These implementations predict the same gauge field content. The 5D action containing the kinetic gauge terms reads
\begin{equation}
S^{\rm 5D}_{\rm gauge}= -\frac{1}{4} \int {\rm d}^5 x \sqrt{g} \, g^{AM} g^{BN} \left( \tr W_{AB}W_{MN} + \tr W^{\prime}_{AB} W^{\prime}_{MN} + B^{\prime}_{AB} B^{\prime}_{MN} \right),
\label{S_5d_gauge}
\end{equation}
with $W$, $W^{\prime}$, and $B^{\prime}$ being the non-abelian 5D gauge field strengths associated to $\mathrm{SU}(2)_L$, $\mathrm{SU}(2)_R$, and $\mathrm{U}(1)_X$, respectively. We denote the corresponding 5D gauge couplings as $g^{\rm 5D}_L$, $g_R^{\rm 5D}$, and $g_X^{\rm 5D}$, whose 4D counterparts are given by $ g_{L,R,X} \equiv g_{L,R,X}^{\rm 5D}/\sqrt{L}$. We did not include the gluon since it does not play a central role in our analysis. The mechanism responsible for the breaking of $\mathrm{SU}(2)_R \times \mathrm{SU}(2)_L \times \mathrm{U}(1)_X$ down to the electroweak (EW) symmetry group, $\mathrm{SU}(2)_L \times \mathrm{U}(1)_Y$, as well as the relations between the various couplings and mixing angles, are described in Ref.~\cite{RScusto,O3}.

In the context of the extended gauge group mentioned in the previous paragraph, the brane-localised Higgs doublet gets promoted to a bi-doublet of $\mathrm{SU}(2)_R \times \mathrm{SU}(2)_L$, uncharged under $\mathrm{U}(1)_X$. When it develops a vacuum expectation value (VEV), the Higgs bi-doublet thus breaks, on the IR brane, together with the 5D boundary conditions, the $\mathrm{SU}(2)_R \times \mathrm{SU}(2)_L\times \mathrm{U}(1)_X$ gauge group down to $\mathrm{U}(1)_{e.m.}$ times a global $\mathrm{SU}(2)_V$, the latter endowing the Higgs sector with a custodial symmetry, which keeps under control the contributions to the $T$ parameter. 

After the usual redefinition the Higgs bi-doublet, $ H \to {\rm e}^{kL} H $, the brane-localised action reads
\begin{equation}
S^{\rm 4D}_{\rm Higgs} = \int {\rm d}^4 x  \left[ \frac{1}{2} \eta^{\mu\nu} \tr D_{\mu} H^{\dagger} D_{\nu} H - \frac{\lambda_0}{4} \left( \tr  H^{\dagger} H - v^2 \right)^2 \right]_{y=L},
\label{S_higgs}
\end{equation}
where $v\simeq 246$~GeV (this is true, as it will be shown in the next subsection, only in the limit where the KK partners decouple). Omitting the gluon, the covariant derivative is given, in terms of the 5D gauge fields, by
\begin{equation}
D_{\mu} H = \partial_{\mu} H - i \sqrt{L} \left[ g_{L} (W_{\mu}^a \, I_L^a) H + g_{R} H (W_{\mu}^{\prime a} \, I_R^a )^T \right],
\label{covar_deriv}
\end{equation}
with $I_{L,R}^a, \, a=1,2,3$ being the $\mathrm{SU}(2)_{L,R}$ generators, proportional to the usual Pauli matrices. The $\sqrt{L}$ factor originates from using 4D couplings instead of 5D (dimensionful) couplings. Besides, due to the scalar bi-doublet having null charge under ${\rm U}(1)_X$ , the $B'$ gauge field does not appear in the covariant derivative acting on $H$.  After EWSB, the Higgs bi-doublet is parametrized as
\begin{equation}
H = \frac{v + h_0 (x)}{\sqrt{2}} \begin{pmatrix} 0 & -1 \\ 1 & 0 \end{pmatrix},
\label{higgs_bidoublet}
\end{equation}
with $h_0$ being the (4D) Higgs field (before mixing with the radion). Putting all these ingredients together, the 4D action has the following expression:
\begin{gather}
S^{\rm 4D}_{\rm Higgs} = \int {\rm d}^4 x \, L  \left( 1 + \frac{h_0}{v} \right)^2 \left[ \bar{m}_W^2 \left( W_{\mu} - \alpha_W W_{\mu}^{\prime} \right)^2 + \frac{\bar{m}_Z^2}{2} \left( Z_{\mu} - \alpha_Z Z_{\mu}^{\prime} \right)^2 \right]_{y=L} \notag \\
+ \int {\rm d}^4 x \left[\frac{1}{2} (\partial_{\mu} h_0)^2-  \left(\frac{m_{h_0}^2}{2} h_0^2 + \frac{m_{h_0}^2}{2 v} h_0^3 + \frac{m_{h_0}^2}{8 v^2} h_0^4 \right) \right].
\label{S_brane}
\end{gather}
Here, $V_{\mu}^2 \equiv \eta^{\mu\nu} V_{\mu} V_{\nu}$, and
$$\alpha_{W}=g_R/g_L, \ \ \alpha_{Z}=\sqrt{g_R^2/g_Z^2-\sin^2 \theta_W}, \ \ g_Z=g_L/\cos \theta_W , $$ 
$\theta_W$ being the weak mixing angle. 
From now on, unless otherwise stated, we will consider the configuration $g_R=g_L$ (as enforced by a Left-Right Parity~\cite{O3}).
The masses in the first line of eq.~\eqref{S_brane} are given by $\bar{m}_{W,Z} = \frac{g_{L,Z} v}{2}$; as we will show in the next section, they are not equal to the measured $W$ and $Z$ boson masses. Moreover, $m_{h_0}^2 =  2 \lambda_0 v^2$ is the Higgs mass in the absence of mixing with the radion. The expression above will be our starting point for deriving the ($y$-dependent) wave functions of the $Z$ boson and its KK partners, as well as their couplings to the mixed Higgs-radion scalar fields.

\subsection{KK Gauge Boson Mixing}

In this subsection, we will outline the procedure employed for obtaining the masses and profiles of the $Z$ boson and its KK partners. We will denote the $Z$ boson by $Z_0$, while its KK excitations (which here are also mass eigenstates) will be referred to as $Z_{n}$, with $n=1$ for the first KK level, $n=2$ for the second one, and so on. Collecting several terms from eqs.~\eqref{S_5d_gauge} and \eqref{S_brane}, the relevant part of the action reads, after EW symmetry breaking, as follows:
\begin{align}
S_{ZZ}^{\rm 5D} &= \int {\rm d}^5 x \sqrt{ g} \left( -\frac{1}{4} \,  g^{AB}  g^{MN} Z_{AM} Z_{BN} - \frac{1}{4} \,  g^{AB}  g^{MN} Z_{AM}^{\prime} Z_{BN}^\prime \right) \notag \\
	& + \int {\rm d}^5 x \,   L \, \delta(y-L) \frac{\bar{m}_Z^2}{2}  \left[Z_{\mu} (x,y) - \alpha_Z Z_{\mu}^\prime (x,y) \right]^2,
\label{ZZ_action}
\end{align}
where $Z_{MN}^{(\prime)} \equiv \partial_M Z_N^{(\prime)} - \partial_N Z_M^{(\prime)}$. We choose to work in a gauge where the fifth component of the 5D gauge fields, $Z_5^{(\prime)}$, is null.~\footnote{While the $0$-mode of the 5D scalar field $Z_5^{(\prime)}$ is set to $0$ by the boundary conditions (BCs), one can interpret the KK modes of $Z_5^{(\prime)}$ as the longitudinal components of the KK Z bosons, $Z^{\mu}_{i\geq 1}$.} Similarly to e.g. Ref.~\cite{Azatov:2010pf}, we will perform a ``mixed'' KK decomposition, but applied to the gauge bosons:
\begin{align}
Z_{\mu} (x,y) &= \frac{1}{\sqrt{L}} \sum_{n\geq 0} g_{+}^n (y) Z_{n,\mu} (x), \notag \\
Z_{\mu}^{\prime} (x,y) &= \frac{1}{\sqrt{L}} \sum_{n\geq 0} g_{-}^n (y) Z_{n,\mu} (x),
\label{KK_decomp_Z}
\end{align}
where the (dimensionless) profiles $g_{\pm}^n$ obey Neumann boundary condition at $y=L$ and Neumann (+) or Dirichlet (-) boundary condition at $y=0$. Choosing (-) boundary conditions at $y=0$ for the $Z^{\prime}$ field eliminates its zero-mode, thus reproducing the low-energy spectrum, made out of a single light $Z$ boson (SM field content). Such a mixed decomposition will allow us to include the boundary-localised mixing between the $Z$ and $Z^{\prime}$ 5D fields into the (coupled) equations of motion for $g_+$ and $g_-$, which in turn will lead us to the exact expressions for the profiles and masses of the KK excitations of the $Z$ boson. 

By using the standard technique of varying the action in eq.~\eqref{ZZ_action} with respect to the $Z_{\mu}$ and $Z_{\mu}^{\prime}$ fields and then employing the KK decomposition in eqs.~\eqref{KK_decomp_Z}, one gets the following equations of motions (EOMs) for the profiles:
\begin{gather}
\partial_5 \left({\rm e}^{-2ky} \partial_5 g_+ \right) + m^2 g_+ = \bar{m}_Z^2 \, L \, \delta(y-L) \left[ g_+ (L) - \alpha_Z \, g_- (L) \right], \notag \\
\partial_5 \left({\rm e}^{-2ky} \partial_5 g_- \right) + m^2 g_- = -\alpha_Z \, \bar{m}_Z^2 \, L  \,\delta(y-L) \left[ g_+ (L) - \alpha_Z \, g_- (L) \right],
\label{EOMs} 
\end{gather}
with the BCs given by
\begin{equation}
g_+^{\prime} (0) = g_{\pm}^{\prime} (L) = g_- (0) = 0,
\label{BCs}
\end{equation}
where the exponent `` $^\prime$ " denotes differentiation with respect to $y$. For better readability, we have suppressed the $n$ indices, which labeled the KK levels. 

\begin{figure}[t]
\begin{center}
\includegraphics[keepaspectratio=true,width=0.45\textwidth]{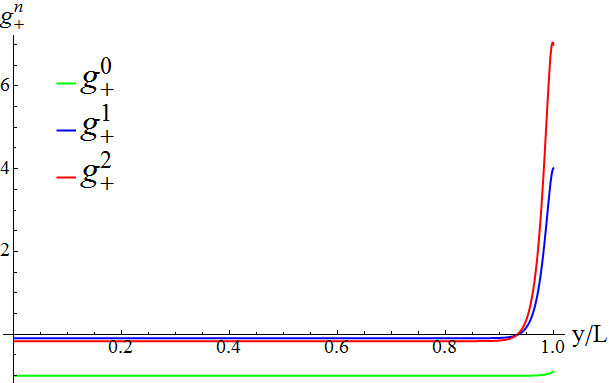}
\quad
\includegraphics[keepaspectratio=true,width=0.45\textwidth]{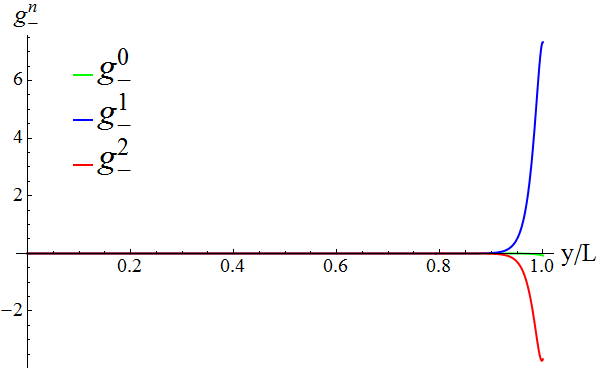}
\caption{
{\small Profiles of the $Z_0$ (green), $Z_1$ (blue), and $Z_2$ (red) fields, corresponding to (left) $(++)$ and (right) $(-+)$ boundary conditions, accordingly to eq.~\eqref{KK_decomp_Z}. 
We have set $m_{KK}=3$~TeV.}  
}
\label{fig:profiles}
\end{center}
\end{figure}

The presence of the delta functions in the EOMs induces discontinuities in the first derivatives of the profiles at $y=L$. To find out by how much the derivatives ``jump'', we integrate the EOMs in eq.~\eqref{EOMs} from $L-\epsilon$ to $L$, and then take $\epsilon\to 0$, which gives us the following relations:
\begin{align}
\bar{m}_Z^2 \, L \, {\rm e}^{-2kL} \left[ g_+ (L) - \alpha_Z g_- (L) \right] + g_+^{\prime} (L_-) &= 0 \notag \\
\alpha_Z \, \bar{m}_Z^2 \, L \, {\rm e}^{-2kL} \left[ g_+ (L) - \alpha_Z g_- (L) \right] - g_-^{\prime} (L_-) &= 0,
\label{jump_cdts}
\end{align}
where we used the notation $\lim_{\epsilon\searrow 0} f(x-\epsilon) \equiv f(x_-)$. We now have all the prerequisites to calculate the profiles and the masses of the $Z$ boson tower. Combining eqs.~\eqref{EOMs}, \eqref{BCs}, and \eqref{jump_cdts}, we find the well-known expressions for the profiles~\cite{exRSflav1}, which are expressed by the Bessel function of the first ($J_{\alpha}$) and second ($Y_{\alpha}$) kinds:
\begin{equation}
g_{\pm}^n = N_{\pm}^n {\rm e}^{ky} \left[ J_{\frac{1}{2} \mp \frac{1}{2}} \left(x_n {\rm e}^{-kL}\right) Y_1 \left(x_n {\rm e}^{k(y-L)}\right) - Y_{\frac{1}{2} \mp \frac{1}{2}} \left(x_n {\rm e}^{-kL}\right) J_1 \left(x_n {\rm e}^{k(y-L)}\right)\right],
\label{Z_profiles}
\end{equation}
where $x_n \equiv 6 \, m_{Z_n} / m_{KK}$. The normalisation constants $N_{\pm}^n$ are obtained by requiring that each $Z_n$ field has a canonically normalised kinetic term, which translates to
\begin{equation}
\int_0^L \frac{{\rm d}y}{L} \, \left( g_+^m g_+^n + g_-^m g_-^n \right) = \delta_{mn}.
\label{norm_cst}
\end{equation}
We plot in Fig.~\ref{fig:profiles} the $(++)$ and $(-+)$ profiles $g_{\pm}^n$ corresponding to the observed $Z$ boson ($n=0$) and to its two lightest KK modes ($n=1,2$). Notice that $g_-^0$ is slightly shifted from $0$ close to $L$ due to the $Z-Z^{\prime}$ mixing. Also, $g_+^0$ is flat in most of the $[0,L]$ interval, with a small departure close to the IR brane, where the mixing of the SM-like $Z$ boson with the heavier KK partners takes place. As for the lowest KK-$Z$ profiles, i.e. $g_{\pm}^{1,2}$, they are all of comparable size and peaked towards $y=L$, signaling the usual KK partner localization close to the IR brane.

Meanwhile, the mass spectrum is obtained by solving the system of equations~\eqref{jump_cdts}. One thus obtains
\begin{equation}
\frac{6 \bar{m}_Z^2 (kL)^2}{m_{KK}^2} \left[ g_+ (L) g_-^{\prime} (L_-) - \alpha_Z^2 \, g_+^{\prime} (L_-) g_- (L) \right] + g_+^{\prime} (L_-) g_-^{\prime} (L_-) = 0.
\label{Z_mass_spectrum}
\end{equation}
Notice that the normalisation constants $N_{\pm}$ simplify in this equation. Since the lightest mode of the $Z$ KK-tower is identified with the observed $Z$-boson, its mass should be equal to the measured $m_Z \simeq 91.2$~GeV. Imposing this condition determines the value of $\bar{m}_Z$ (and thus, as discussed later, of $v$) as a function of the mass of the first KK excitation of the photon/gluon, $m_{KK}$. In turn, knowing $\bar{m}_Z$, one can compute the masses of the KK eigenstates associated to the $Z$ boson. 

\begin{figure}[h]
\begin{center}
\includegraphics[keepaspectratio=true,width=0.55\textwidth]{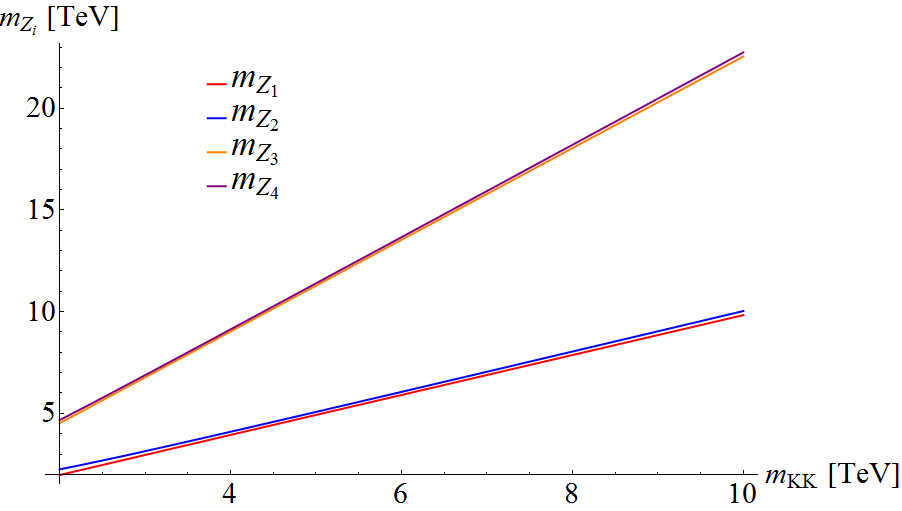}
\caption{
{\small Masses (in TeV) of the first four KK $Z$ boson eigenstates, as a function of the first KK photon mass, $m_{KK}$ (in TeV).}  
}
\label{fig:kk_masses}
\end{center}
\end{figure}

We display in Fig.~\ref{fig:kk_masses} the first four KK $Z$ mass eigenvalues as a function of the KK photon mass, $m_{KK}$. As expected, $m_{Z_1}$ and $m_{Z_2}$ are almost degenerate and of the order $m_{KK}$ (the first $Z'$ mode mass is close to $m_{KK}$), with a mass splitting of $\sim 100-200$~GeV, i.e. of the order of the electroweak scale (order of the off-diagonal mixing mass term). In the limit of zero mixing, $Z_1$ ($Z_2$) would correspond to the first KK mode of the $Z^{\prime}$ ($Z$) gauge boson. For similar reasons, $m_{Z_3}$ and $m_{Z_4}$ are nearly degenerate at a scale
such that $m_{Z_3}-m_{Z_1}$ is much larger than the electroweak scale.

\vspace{0.5cm}

In fact, $\bar{m}_Z$ quantifies nothing else than the Higgs doublet VEV shift~\cite{BouchartZZH}. This phenomenon arises from the fact that the $Z$ boson does not acquire its mass only from the scalar VEV, but also from mixing with the heavier KK partners. Therefore, to reproduce the very precisely measured $m_Z$, the VEV should be adjusted. To first non-trivial order in $m_Z / m_{KK}$, the RS VEV $v$ gets shifted from its SM value $v_{SM}$ as
\begin{align}
v \simeq v_{SM} \left( 1 + \frac{1+\alpha_Z^2}{2} \frac{3 m_Z^2 kL}{m_{KK}^2} \right).
\label{vev_shift}
\end{align}
There are also other contributions at order $m_Z^2/m_{KK}^2$, but we do not display them, as they are not enhanced by the so-called volume factor, $kL$. Nevertheless, in our calculations we will use the exact value of the obtained shifted VEV, $v$. As eq.~\eqref{vev_shift} already shows, the shifted VEV is always larger the SM VEV, i.e. $v>v_{SM}=246$~GeV (in the decoupling limit $m_{KK}\to\infty$, the two VEVs become equal, as expected).

For later use in the expression of the $\phi Z$ and $hZ$ cross sections, we also give the couplings of the $Z_i$ eigenstates to the light fermions which constitute the initial state for the process we are considering ($e^{\pm}$ for the ILC and light quark flavours, $u,d,s,c$, for the LHC). Since we will consider the main intermediate states exchanged in the s-channel, that is only the first $Z_{i=0,1,2}$
states, i.e. the $Z$ boson and its first two excitations, we will consider only their couplings to the light fermions. Such couplings can be inferred from the covariant derivative of the 4D part of the kinetic term of the 5D fermionic field:
\begin{equation}
S_{\Psi}^{\rm 5D} = \int {\rm d}^5 x \sqrt{ g} \, \bar{\Psi} \, i \Gamma^{\mu} D_{\mu} \Psi \to \int {\rm d}^5 x \sqrt{ g} \, \sqrt{L} \, \bar{\Psi} \, \Gamma^{\mu} (g_Z Q_Z^{\Psi} Z_{\mu} + g_{Z^{\prime}} Q^{\Psi}_{Z^{\prime}} Z^{\prime}_{\mu})  \Psi,
\label{ferm_kin_term}
\end{equation}
where $\Psi$ denotes a generic 5D fermion, whose zero mode is a light SM fermion. The $\sqrt{L}$ factor allows us to use the 4D couplings $g_Z$ (defined in the previous subsections) and $g_{Z^{\prime}}= g_R/\sqrt{g_R^2 - g_L^2 \tan^2 \theta_W}$, instead of their (dimensionful) 5D equivalents. Meanwhile, $Q^{\Psi}_{Z^{(\prime)}}$ is the $Z^{(\prime)}$ charge of the fermion $\Psi$, given by 
\begin{equation}
Q_Z^{\Psi} = I^{\Psi}_{3L} - Q_{\gamma}^{\Psi} \sin^2 \theta_W, \quad Q^{\Psi}_{Z^{\prime}} =I^{\Psi}_{3R} - Y^{\Psi} \frac{g_L^2 \tan^2 \theta_W}{g_R^2},
\label{Z_charges}
\end{equation}
with $I^{\Psi}_{3L/R},Q^{\Psi}_{\gamma},Y^{\Psi}$ being, respectively, the left/right isospin quantum number, electric charge and hypercharge of the fermion $\Psi$. Denoting by $\exp(3ky/2) \, f(y)$ the profile of the light SM fermion originating from $\Psi$, one obtains its couplings to the $Z_i$ bosons by plugging the KK decomposition in eq.~\eqref{KK_decomp_Z} into eq.~\eqref{ferm_kin_term}, thus obtaining
\begin{equation}
g_Z Q_Z^{\Psi} \int_0^L \frac{{\rm d}y}{L} \, f^2 (y) \, g_+^i (y) + g_{Z^{\prime}} Q^{\Psi}_{Z^{\prime}} \int_0^L \frac{{\rm d}y}{L} \, f^2(y) \, g_-^i (y) \equiv g_Z Q_Z^{\Psi} c_i.
\label{ferm_Z_coupl1}
\end{equation}
These couplings can easily be deduced from profile overlap considerations. First, note that the light fermion profiles, which will be relevant for the initial state particles, are peaked towards the UV brane, with very small values close to the IR brane. Meanwhile, as shown in Fig.~\ref{fig:profiles}, the $g_{\pm}^{i=0,1,2}$ profiles are almost constant along the extra dimension, the sole exception being a small region near the IR brane, where they get peaked. Consequently, the overlap between the $g_{\pm}$'s and the light fermion profiles will effectively take place only in a region close to the UV brane, where the gauge boson profiles are almost constant. Therefore, bearing in mind that the fermion profiles are orthonormalised, the overlap between the light fermionic profiles and the gauge boson wave function are excellently approximated by the simple expression
\begin{equation}
c_i \simeq g_+^i (0) \int_0^L \frac{{\rm d}y}{L} \, f^2 (y) =  g_+^i (0).
\label{ferm_Z_coupl2}
\end{equation}
The $g_-$ profiles do not appear in this expression simply because their boundary conditions imply $g_-^i (0) = 0$. Therefore, in some sense, the light fermions couple only to the $\rm{SU}(2)_L \times \rm{U}(1)_Y$ ``part'' of the $Z$ KK tower, which means that only their (SM-like) representations under the aforementioned gauge group will be relevant for their coupling to the $Z_{0,1,2}$ states.

\subsection{Higgs and Radion Couplings Before Mixing}  \label{sec:HRcoupl}

We now focus on the radion and how it couples to the $Z$ boson KK tower. We start by taking the background RS metric from eq.~\eqref{static_metric} and including the scalar perturbation $F(x,y)$ 
as in Ref.~\cite{RadHatIR},
\begin{equation}
{\rm d} s^2 = {\rm e}^{-2(k\,y+F)} \eta_{\mu\nu} {\rm d}x^{\mu}{\rm d}x^{\nu} - (1+2F)^2 {\rm d}y^2 \equiv \bar g_{MN} {\rm d}x^{M}{\rm d}x^{N},
\label{perturbed_metric}
\end{equation}
where we used $\bar g_{MN}$ to denote the 5D metric with scalar perturbations included, in order to differentiate it from its unperturbed counterpart, $g_{MN}$. The linearized metric perturbations read
\begin{equation}
\bar g_{MN} - g_{MN} \equiv \delta g_{MN} \simeq -2F \, {\rm diag} \left({\rm e}^{-2 k\,y} \eta_{\mu\nu},2\right).
\label{linear_metric_pert}
\end{equation}
The situation is slightly different for terms localised on the IR brane, i.e. terms that contain the Higgs bi-doublet. On this brane, the line element is written as
\begin{equation}
{\rm d} s^2_{\rm IR} = {\rm e}^{-2[kL + F(x,L)]} \eta_{\mu\nu} {\rm d}x^{\mu}{\rm d}x^{\nu} \to {\rm e}^{-2 F(x,L)} \eta_{\mu\nu} {\rm d}x^{\mu}{\rm d}x^{\nu} \equiv \bar \eta_{\mu\nu} {\rm d}x^{\mu}{\rm d}x^{\nu},
\label{perturbed_metric_IR}
\end{equation}
where the arrow was used to indicate that the redefinition of the Higgs bidoublet $H$ absorbs away the ${\rm e}^{-2kL}$ factor. Therefore, the linearized metric perturbations on the IR brane are given by
\begin{equation}
\bar{\eta}_{\mu\nu} - \eta_{\mu\nu} \equiv \delta \eta_{\mu\nu} \simeq -2 F(x,L) \eta_{\mu\nu}.
\label{linear_metric_pert_IR}
\end{equation}
In the limit of small backreaction (of the field $F$ on the metric curvature), 
the scalar perturbation $F(x,y)$ can be parametrized as follows~\cite{RadHatIR}:
\begin{equation}
F(x,y) = \frac{\phi_0 (x)}{\Lambda} {\rm e}^{2 k (y-L)},
\label{radion_profile}
\end{equation} where $\phi_0$ is the (unmixed) 4D radion field~\footnote{The KK radion modes are absorbed into the (longitudinal) degrees of freedom of the massive KK gravitons.} 
and $\Lambda$ is the radion VEV, which is an $\mathcal{O}$~(TeV) energy scale that sets the length of the extra dimension~\cite{GW}. 
At linear order, the radion's interaction with the gauge fields and the Higgs can be obtained by making the following replacements:
\begin{itemize}
\item  $g^{MN} \to \bar g^{MN}$ in eq.~\eqref{S_5d_gauge} for interactions originating from the bulk terms, 
\item ${\rm d}^4 x \to {\rm d}^4 x  \sqrt{\bar \eta}$, $\eta^{\mu \nu} \to\bar \eta^{\mu \nu}$ in eq.~\eqref{S_brane} for brane-localised interactions,
\end{itemize}
and then keep only the terms linear in $F$.~\footnote{Equivalently, one can find the radion couplings by varying the action with respect to the metric and keeping only the linear metric perturbations~\cite{RadHatIR}.} Finally, to derive the effective 4D couplings, and take into account the KK Z mixings, 
one should employ the KK expansion from eq.~\eqref{KK_decomp_Z} and perform the usual integration over $y$ (or, for the brane-localised terms, just evaluate the profiles in $y=L$). Thus, putting all these elements together, we arrive at the complete 4D Lagrangian describing the $h_0 ZZ$ and $\phi_0 ZZ$ interactions:
\begin{equation}
\mathcal{L}_{\varphi Z_i Z_j}^{\rm 4D} = \bar{m}_Z^2 \left( \frac{h_0}{v} - \frac{\phi_0}{\Lambda
} \right) C_i^{\rm 4D} C_j^{\rm 4D}  Z_{i,\mu} Z_j^{\mu}
-\frac{\phi_0}{\Lambda}\left[ \frac{m_{KK}^2}{3 (kL)^2} \, C^{5D}_{ij} Z_{i,\mu} Z_j^{\mu} + \frac{1}{2}\tilde{C}_{ij}^{\rm 5D} Z_{i,\mu \nu} Z_j^{\mu\nu} \right],
\label{L_ZZ_scalar}
\end{equation} 
where we have used the following notations:
\begin{align}
C_i^{\rm 4D} &= g_+^i (L) - \alpha_Z \, g_-^i (L), \label{c4} \\
C_{ij}^{\rm 5D} &= L \int_0^L {\rm d} y \left[ (g_+^i)^\prime (g_+^j)^\prime + (g_-^i)^\prime (g_-^j)^\prime  \right], \label{c5} \\
\tilde{C}_{ij}^{\rm 5D} &= \frac{1}{L} \int_0^L {\rm d} y \, {\rm e}^{2 k(y-L)} \left( g_+^i g_+^j + g_-^i g_-^j \right). \label{c5_tilde}
\end{align}
Let us now trace the origin of each term appearing in eq.~\eqref{L_ZZ_scalar}. The first term, proportional to $\bar{m}_Z^2$, originates from the brane-localised mass term in the first line of eq.~\eqref{S_brane}, whereas the terms between square brackets come from the 5D gauge kinetic terms in eq.~\eqref{S_5d_gauge}. More precisely, in terms of 5D fields, the first term between the square brackets originates from the $Z_{5\mu} Z^{5\mu}$ term, while the second one stems from $Z_{\mu\nu} Z^{\mu\nu}$.

We now have all the ingredients to derive the mixed Higgs-radion couplings to the $Z_{i}$ bosons, which we will do in the next section.

\subsection{Higgs-Radion Mixing and Couplings}

The Higgs-radion mixing arises at the renormalisable level by coupling the 4D Ricci scalar $R_4$ to the trace of $H^{\dagger} H$
via a possible gauge invariant term~\cite{RadSMatIR} as follows:
\begin{equation}
S_{\xi}^{\rm 4D} = \xi \int {\rm d}^4 x \, \sqrt{\bar{\eta}} \, R_{4 }(\bar \eta_{\mu\nu}) \frac{1}{2} \tr \left( H^{\dagger} H  \right),  
\label{hr_mixing}
\end{equation}
with $\bar \eta_{\mu\nu}$, the perturbed IR brane metric, defined in eq.~\eqref{perturbed_metric_IR}. As it involves the brane-localised Higgs field, the Higgs-radion mixing comes from the IR brane. A non-zero $\xi$ coupling in eq.~\eqref{hr_mixing} induces a kinetic mixing between the two scalars after EW symmetry breaking, the Higgs-radion Lagrangian at the quadratic level being given by~\cite{Frank,RadHatIR,RadSMatIR}
\begin{equation}
\mathcal{L}_{\varphi\varphi}^{\rm 4D} = -\frac{1}{2} \begin{pmatrix} \phi_0 & h_0 \end{pmatrix} \begin{pmatrix}
1 + 6 \xi \ell^2 & -3 \xi \ell \\ -3 \xi \ell & 1  \end{pmatrix}
\begin{pmatrix} \Box \phi_0 \\ \Box h_0 \end{pmatrix} - \frac{1}{2} m_{\phi_0}^2 \phi_0^2 - \frac{1}{2} m_{h_0}^2 h_0^2,
\label{lagr_mixing}
\end{equation}
where $\ell \equiv v / \Lambda$ is the ratio between the Higgs and radion VEVs and $\Box$ is the flat-space d'Alembertian. 
The transition to the mass eigenstates, $\phi$ and $h$, is achieved through a non-unitary transformation
diagonalising the kinetic terms of eq.~\eqref{lagr_mixing}:
\begin{equation}
\begin{pmatrix} \phi_0 \\ h_0 \end{pmatrix} = \begin{pmatrix} a & -b \\ c & d \end{pmatrix} \begin{pmatrix} \phi \\ h \end{pmatrix}.
\label{rot_matrix}
\end{equation}
Using notations similar to the ones in Ref.~\cite{RadHatIR}, the elements of this matrix are $a = \cos\theta / Z$, $b=\sin \theta /Z$, $c= \sin \theta + t \cos \theta$, and $d=\cos\theta - t \sin \theta$, with $t= 6\xi\ell/Z$ and $Z^2 = 1 + 6\xi\ell^2(1-6\xi)$ being the determinant of the kinetic mixing matrix from eq.~\eqref{lagr_mixing}. The mixing angle is given by
\begin{equation}
\tan \theta = \frac{m_{h_0}^2-m_h^2}{t \, m_{h_0}^2} = -\frac{t \, m_{h_0}^2}{m_{h_0}^2-m_{\phi}^2} .
\label{theta_scalar}
\end{equation}
The squared mass $m_{h_0}^2$ can then be expressed in terms of the physical mass eigenvalues $m_{h,\phi}$ as follows~\cite{RadSMatIR}:
\begin{equation}
m_{h_0}^2 = \frac{Z^2}{2} \left[ m_h^2 + m_{\phi}^2 + {\rm sign}(m_h^2 - m_{\phi}^2) \sqrt{(m_h^2 - m_{\phi}^2)^2-\frac{144 \, \xi^2 \ell^2 m_h^2 m_{\phi}^2}{Z^2}} \right],
\label{m_h0}
\end{equation}
while $m_{\phi_0}^2$ can be deduced from $ m_{h_0}^2 m_{\phi_0}^2 = Z^2 m_h^2 m_{\phi}^2 $, which results from evaluating the mass matrix determinant in both bases. As it is clear from the expression of $m_{h_0}^2$ above, we use the sign convention in which $m_h$ ($m_{\phi}$) coincides with $m_{h_0}$ ($m_{\phi_0}$) when $\xi=0$.

Summing up, the Higgs-radion system is described by four parameters: the mixing parameter $\xi$, the radion VEV $\Lambda$, the physical radion mass $m_{\phi}$, and the physical Higgs mass $m_h$, which we fix at $125$~GeV. There is also a fifth parameter, the first KK photon mass $m_{KK}$, which enters indirectly into this interplay by shifting the Higgs VEV. However, one cannot take arbitrary values for these parameters, as there are two theoretical consistency conditions which constrain the parameter space. The first condition is the absence of ghost fields in the theory, which restricts the kinetic mixing matrix determinant to positive values, i.e. $Z^2>0$. The second one concerns the square root appearing in eq.~\eqref{m_h0}, whose argument should be positive. This gives the following mathematical condition:
\begin{equation}
Z^2 (m_h^2 - m_{\phi}^2)^2 \geq 144 \, \xi^2 \ell^2 m_h^2 m_{\phi}^2,
\label{th_excl}
\end{equation}
which actually supersedes the no-ghost condition, $Z^2>0$, in the whole parameter space. Note that, in the case of exact degeneracy between the Higgs and the radion, there can be no Higgs-radion mixing, as the condition in eq.~\eqref{th_excl} imposes $\xi = 0$ if $m_h=m_{\phi}$.

We can now express the couplings of the physical Higgs and radion states to the gauge bosons. To ease the notations, we will use the following definitions, which are similiar to the ones in Ref.~\cite{RadHatIR}:
\begin{equation}
g_{\phi}=c-\ell a, \quad g_h = d + \ell b, \quad g_{\phi}^r =-\ell a, \quad g_h^r = \ell b.
\label{g_phi}
\end{equation}
Using these definitions and the couplings of $\phi_0, h_0$, which were derived in the previous section, one can straightforwardly write down the couplings for the scalar mass eigenstates, $\phi$ and $h$. As we are focusing on the $Z\phi$ (and $Zh$) production mechanism, we first list the Lagrangian for $\phi Z_i Z_j$ interactions, which is obtained by inserting the definitions of eq.~\eqref{g_phi} in eq.~\eqref{L_ZZ_scalar}:
\begin{align}
\mathcal{L}_{\phi Z_i Z_j}^{\rm 4D} &= \frac{\bar{m}_Z^2}{v} \left( g_{\phi} \, C_i^{\rm 4D} C_j^{\rm 4D} +  \frac{ g_{\phi}^r \, m_{KK}^2}{3 \bar{m}_Z^2 (kL)^2} C^{5D}_{ij} \right)  \phi Z_{i,\mu} Z_j^{\mu} + \frac{g_{\phi}^r}{2 v} \tilde{C}_{ij}^{\rm 5D} \phi Z_{i,\mu \nu} Z_j^{\mu\nu} \notag \\
&\equiv \frac{\bar{m}_Z^2}{v} \phi \left[ C_{ij}^{\phi} \,   Z_{i,\mu} Z_j^{\mu} + \frac{\tilde C_{ij}^{\phi}}{2\, \bar{m}_Z^2} \,  Z_{i,\mu \nu} Z_j^{\mu\nu} \right] .
\label{phi_ZZ}
\end{align}
The $hZ_i Z_j$ interactions are obtained by simply substituting $\phi\to h$ in the above equation.

\begin{figure}[t]
\begin{center}
\includegraphics[keepaspectratio=true,width=0.43\textwidth]{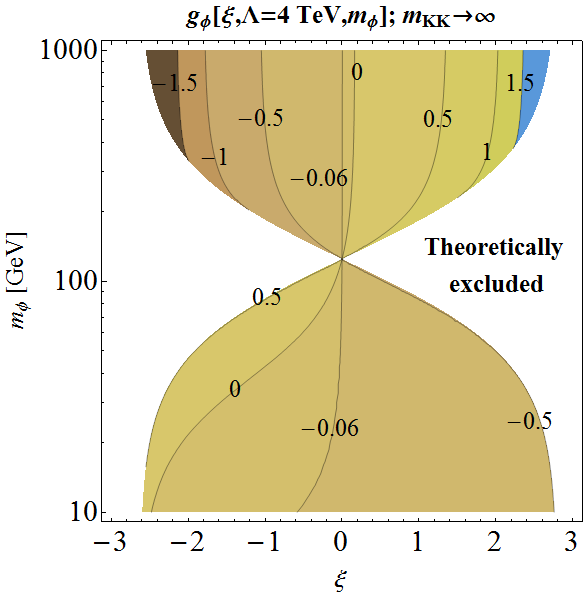}
\quad
\includegraphics[keepaspectratio=true,width=0.43\textwidth]{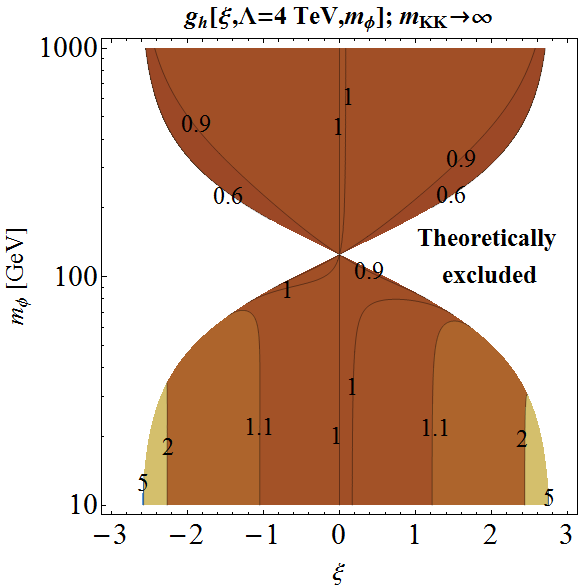}
\includegraphics[keepaspectratio=true,width=0.43\textwidth]{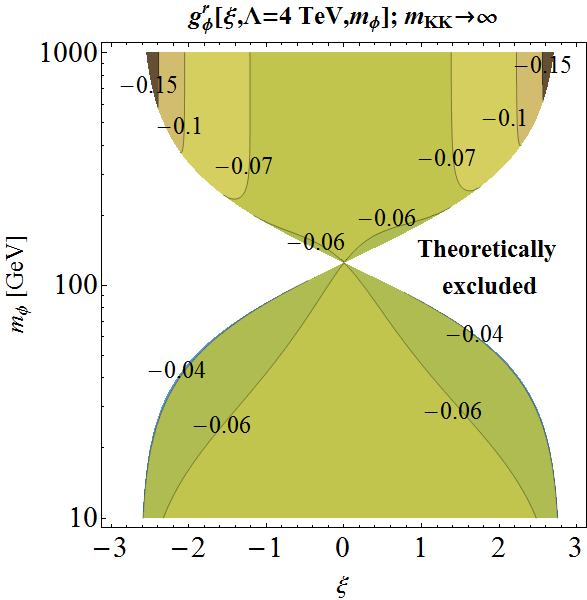}
\quad
\includegraphics[keepaspectratio=true,width=0.43\textwidth]{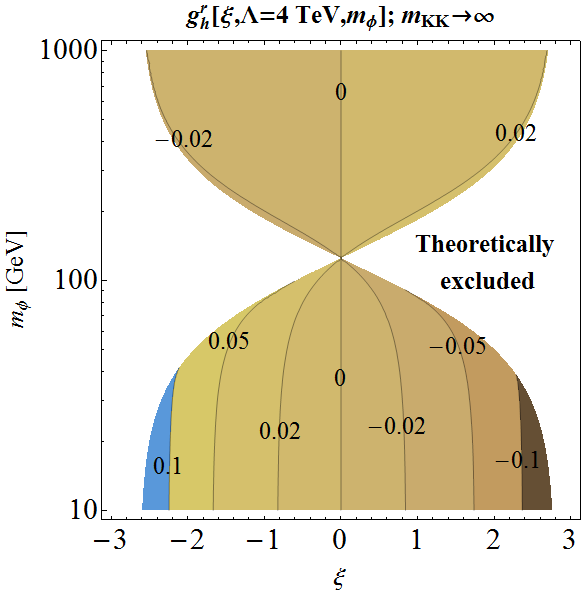}
\caption{
{\small Iso-contours of the couplings (upper left) $g_{\phi}$, (upper right) $g_h$, (lower left) $g_{\phi}^r$, and (lower right) $g_h^r$, in the $\{\xi,m_{\phi}\}$ plane. The four dimensionless couplings plotted above are defined in eq.~\eqref{g_phi}. The white region is excluded by the theoretical consistency condition displayed in eq.~\eqref{th_excl}. The radion VEV $\Lambda$ has been fixed at $4$~TeV, while we have taken, for simplicity, $m_{KK}\to\infty$.}
}
\label{fig:g_phi}
\end{center}
\end{figure}

We plot in Fig.~\ref{fig:g_phi}, as a function of $\xi$ and $m_{\phi}$, the four couplings defined in eq.~\eqref{g_phi}, namely $g_{\phi,h}$ and $g_{\phi,h}^r$. We have chosen $\Lambda=4$~TeV, and, for simplicity, $m_{KK}\to\infty$. In fact, a finite $m_{KK}$ would produce a shift in $v$ and, as the four couplings depend on $\Lambda$ only through the combination $\ell = v/\Lambda$, such a VEV shift can be compensated by adjusting $\Lambda$ to give the same $\ell$. Hence, the value of $m_{KK}$ is not crucial in this context, which is why we have set it to infinity. As the four plots indicate, in most of the parameter space $g_{\phi,h}$ dominates over the $g_{\phi,h}^r$ coupling values. In practice, at currently accessible collider energies, one can ignore the $g_{\phi,h}^r$ couplings when calculating the $Z\phi$ or $Zh$ production cross section (even if those coupling contributions are included in our numerical calculations). 
An exception to this rule applies in the vicinity of the $g_{\phi}=0$ contour~\footnote{At high enough $m_{\phi}$, the $g_{\phi}=0$ condition becomes equivalent to the so-called conformal limit, $\xi=1/6$.}: in this region, $g_{\phi}^r$ becomes dominant, and the radion's coupling to a pair of $Z$ bosons is dramatically reduced, as is the $Z\phi$ production cross section, which tends to render this region blind to current hadronic or even future leptonic colliders. To conclude on this figure, in the limit of KK decoupling (where $C_0^{\rm 4D}\to 1$), the radion coupling to two Z bosons corresponds mainly to $g_{\phi}$ [dimensionless with the normalisation of eq.~\eqref{phi_ZZ}] and is thus driven by the Higgs-radion mixing [see eq.~\eqref{g_phi}].

Before closing this section, let us a remark on the correlation between the first KK photon/gluon mass, $m_{KK}$, and the radion VEV, $\Lambda$. The two quantities are related in the following way:
\begin{equation}
\frac{m_{KK}}{\Lambda} \simeq \frac{k}{M_{\rm Pl}} \ ,
\label{lambda_limit}
\end{equation}
$M_{\rm Pl}$ being the Planck mass.
In order to avoid significant 5D quantum gravitational corrections, the above ratio should satisfy $k/M_{\rm Pl} \lesssim 3$~\cite{Agashe:2007zd}. Throughout
the paper we indeed systematically consider $m_{KK}$ to be smaller than $3\,\Lambda$. Even when the $m_{KK} \to \infty$ limit is considered, it means in fact
that the KK partners are sufficiently heavy so as to not influence the numerical results, i.e. $m_{KK}=\mathcal{O}(10)$~TeV. Such values of $m_{KK}$ do not conflict with 
the considered values of $\Lambda=4,5$~TeV.

\section{The $\phi Z$ and $h Z$ Production} \label{sec:RZ}

We now turn to the study of the $\phi Z/h Z$ production at the LHC and at the ILC, which proceeds through the s-channel exchange of $Z_i$ bosons, $q\bar{q} / e^+ e^- \to Z_i \to Z_0 \phi /Z_0 h$. As higher KK levels are to a very good approximation decoupled, we will only consider the $Z$ boson plus its first two KK excitations, i.e. $i=0,1,2$, as intermediate s-channel states. Moreover, in the LHC case, we consider only the dominant first and second generation quarks as initial state partons. The Feynman rule for the $Z_i Z_0 \phi$ vertex can be straightforwardly deduced from the Lagrangian piece in eq.~\eqref{phi_ZZ}. We display below the squared absolute value of the spin-averaged and polarisation-summed Lorentz invariant amplitude:
\begin{gather}
\overline{\left| \mathcal{M}_{\phi Z} \right|^2} = \frac{g_Z^4 (v_f^2+a_f^2)}{8}  \sum_{i,j=0}^2 \frac{c_i c_j s^2}{(s-m_{Z_i}^2 + {\rm i} \, m_{Z_i} \Gamma_{Z_j})(s-m_{Z_j}^2 - {\rm i} \, m_{Z_j} \Gamma_{Z_j})} \times \notag \\
\times \left[ \frac{\bar{m}_Z^2}{ m_Z^2} (\lambda \sin^2 \theta^{*} + 4 r_Z ) (C_{ij}^{\phi})^2 + 8 \sqrt{\lambda + 4 r_Z} \, C_{ij}^{\phi} \tilde{C}_{ij}^{\phi} + \frac{s}{\bar{m}_Z^2} \left( \lambda (1+\cos^2 \theta^*) + 12 r_Z \right) (\tilde{C}_{ij}^{\phi})^2 \right] ,
\label{amplitude}
\end{gather}
where the coupling factors $c_i$ are defined in eq.~\eqref{ferm_Z_coupl2}.
The notations we used are as follows: $v_f$ and $a_f$ are, respectively, the vectorial and axial couplings of the initial state fermions to the $Z$ boson (i.e. $v_f = I^f_{3L} /2 - Q_{\gamma}^f \sin^2 \theta_W$ and $ a_f = I^f_{3L} /2 $, with $I^f_{3L}$ the weak isospin of the fermion $f$, and $Q_{\gamma}^f$ its electric charge), $\sqrt{s}$ the $e^+ e^-$/partonic center-of-mass energy, and $\theta^*$ the scattering angle in the center-of-mass frame. Moreover, $\lambda = (1-r_{\phi}-r_Z)^2 - 4 r_{\phi}r_Z$, with $r_A = m_A^2/s$, is the usual 2-body phase space function. The wave function overlap integrals $c_i$ were defined previously in eq.~\eqref{ferm_Z_coupl2}. As before, the amplitude for the $Zh$ production process is obtained trivially from eq.~\eqref{amplitude} by changing $\phi \to h$. The expression of the $\phi Z/hZ$ production cross section (in the case of LHC, at the partonic level) is obtained from the integration over $\cos \theta^*$ of the amplitude displayed in eq.~\eqref{amplitude}. 

As it is customary, we denote by $\Gamma_{Z_i}$ the widths of the observed $Z$ boson ($i=0$) and of its first two KK excitations ($i=1,2$). In our calculations, as the (partonic) center-of-mass energy is always above $m_{Z_0}$, we can safely neglect $\Gamma_{Z_0}$. Regarding $Z_{1,2}$, their widths are approximately equal to $10\%$ of their masses. For example, if one takes $m_{KK}=3$~TeV, we get
\begin{equation}
m_{Z_1} \simeq 2.96~{\rm TeV}, \; \Gamma_{Z_1} \simeq 270~{\rm GeV} \quad {\rm and} \quad m_{Z_2} \simeq 3.15~{\rm TeV}, \; \Gamma_{Z_2} \simeq 300~{\rm GeV},
\end{equation}
where we have chosen the dimensionless bulk mass parameters of the top and bottom quarks to be $c_{Q_L}=0.4$, $c_{t_R}=0$, and $c_{b_R}=-0.57$, such that their measured masses are reproduced and the left and right $Zbb$ couplings are close to their SM values. These are values of the $c$-parameters that we will use in our analysis. On the other hand, in order to explain the anomaly on the bottom quark forward-backward asymmetry $A_{\rm FB}^b$ at LEP (and, to a lesser extent,  the anomalous top quark asymmetry $A_{\rm FB}^t$ measured at Tevatron), a more suitable choice would be $c_{Q_L}=0.51$, $c_{t_R}=-1.3$, and $c_{b_R}=0.53$~\cite{RSAFBb,RSAFBt}. In this case, the widths of the KK $Z$ partners change, but not dramatically: $\Gamma_{Z_1} \simeq 350~{\rm GeV}$ and $\Gamma_{Z_2} \simeq 275~{\rm GeV}$. In both cases mentioned above, the Higgs-radion parameters have been fixed as follows: $\xi=1$, $\Lambda=4$~TeV, and $m_{\phi}=750$~GeV. However, the width dependence on these parameters is weak, as the decay to $Z\phi$ is always subdominant. Throughout most of the parameter space spanned by $\xi$, $\Lambda$, and $m_\phi$, with the $c$-parameters chosen above, the dominant decay channel for $Z_1$ ($Z_2$) is to $WW$ ($Zh$).

\subsection{At the LHC} \label{sec:AtLHC}

\subsubsection{Radion Production} \label{sec:AtLHCRad}

The LHC cross-section is obtained by convoluting the cross section for the hard scattering, $\sigma (q\bar{q} \to Z_i \to Z\phi)$, with the parton distribution functions (PDFs). In the following, we use the MSTW set of PDFs at NNLO~\cite{Martin:2009iq}.

\begin{figure}[t]
\begin{center}
\includegraphics[keepaspectratio=true,width=0.55\textwidth]{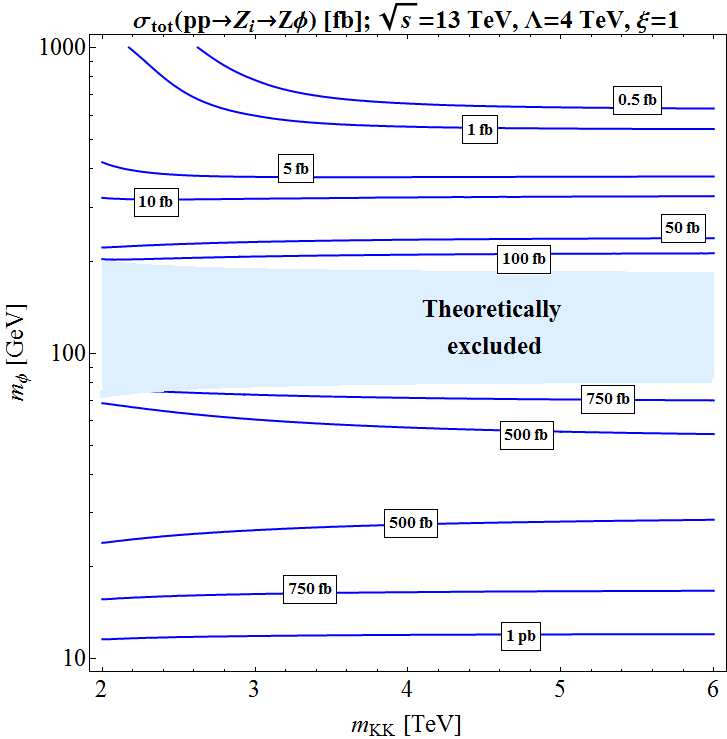}
\caption{
{\small Contour lines of the $Z\phi$ production cross section (in fb and pb) at the LHC in the plane $m_{\phi}$ (in GeV) versus $m_{KK}$ (in TeV). The values of the other involved parameters are $\xi=1$ and $\Lambda=4$~TeV. The light blue region is excluded by the theoretical constraint from eq.~\eqref{th_excl}.}  
}
\label{fig:LHC_phiZ_mkk}
\end{center}
\end{figure}

We first show in Fig.~\ref{fig:LHC_phiZ_mkk} the $Z\phi$ production cross section as a function of $m_{\phi} $ and $m_{KK}$, for a proton-proton center-of-mass energy of $\sqrt{s}=13$~TeV, with $\xi=1$ and $\Lambda=4$~TeV. We consider $m_{KK}$ values above $\sim 2$~TeV as allowed from the direct $Zh$ searches at LHC (potentially affected by  
KK Z mixings), since there is no specific reason to expect unknown effects in this tree-level production -- as discussed in the introduction.
The radion mass range was discussed as well in Section~\ref{se:intro}.

For $m_{KK}\gtrsim 5$~TeV, we see on Fig.~\ref{fig:LHC_phiZ_mkk} that the KK partners of the $Z$ boson no longer play a significant role in the $Z\phi$ production, thus effectively decoupling. This is due to the fact that, at partonic center-of-mass energies $\sqrt{\hat{s}}$ bigger than $\sim 5$ TeV, or equivalently $\hat{s}/s \equiv \tau  \gtrsim (5/13)^2$, the quark-anti-quark luminosity drops down to a negligible level which restricts the on-shell production of $Z_{1,2}$ states. On the contrary, for $m_{KK} \lesssim 5$~TeV, the $Z_1$ and $Z_2$ states play an important role, but only for a radion heavier than $\sim 500$~GeV. This is because, in order to produce a radion plus a $Z$ boson, $\sqrt{\hat{s}}$ should surpass $m_{\phi}+m_Z$, which means that, for a $500$~GeV radion, the virtual $Z$ boson contribution to the $Z\phi$ production is cut off by the $\sqrt{\hat{s}}$ threshold and hence 
becomes comparable to the contribution of its KK partners, $Z_1$ and $Z_2$. However, as one goes to lower radion masses, the cross section dependence on $m_{KK}$ becomes less and less important, as the exchanged virtual $Z$ boson becomes less and less off-shell and starts to dominate over the contributions coming from the exchanges of $Z_1$ and $Z_2$. Nevertheless, we observe a small dependence on $m_{KK}$ for small radion masses as well: its origin lies in the dependence of the $\phi Z_0 Z_0$ coupling on $m_{KK}$, which is a result of the mixing of the SM-like $Z$ boson with its KK partners.

\begin{figure}[t]
\begin{center}
\includegraphics[keepaspectratio=true,width=0.45\textwidth]{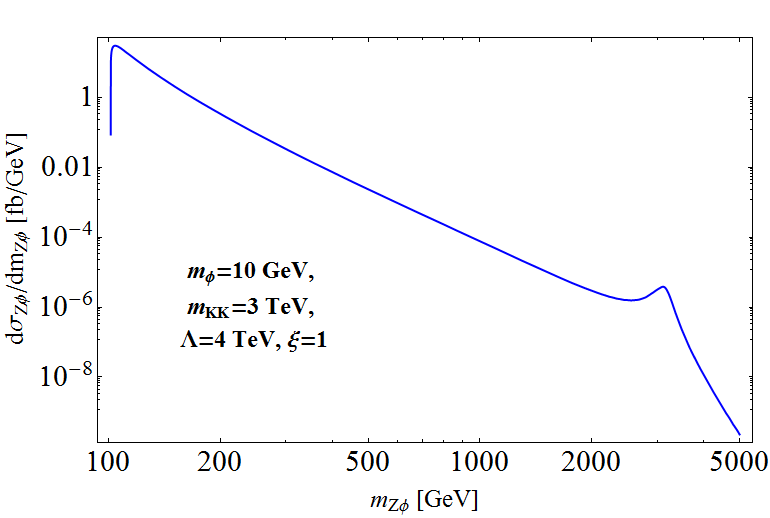}
\quad
\includegraphics[keepaspectratio=true,width=0.45\textwidth]{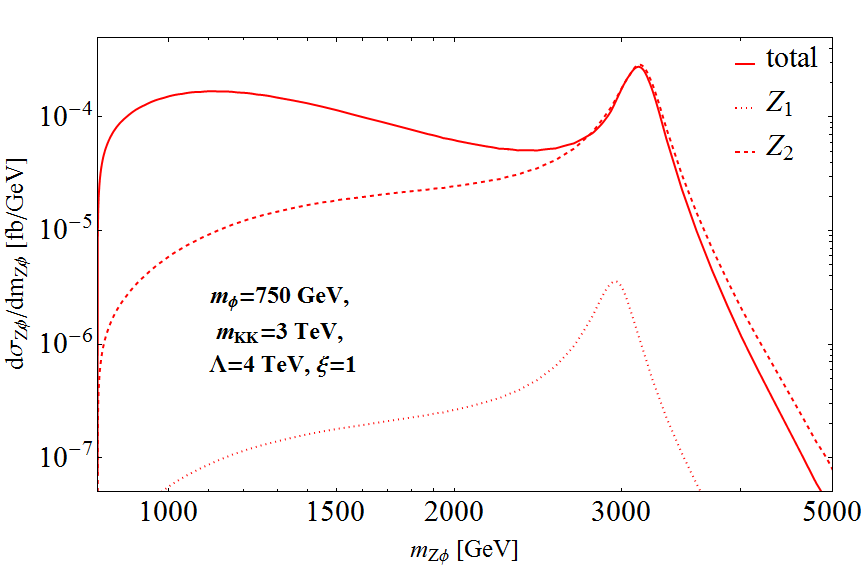}
\caption{
{\small $Z\phi$ invariant mass distribution at LHC (in fb/GeV) for (left) $m_{\phi} = 10$~GeV and (right) $m_{\phi} = 750$~GeV. The other parameters are fixed as follows: $m_{KK} = 3$~TeV, $\Lambda = 4$~TeV, and $\xi=1$. On the right plot, we also display the individual contributions from the two KK boson eigenstates, $Z_1$ and $Z_2$.}  
}
\label{fig:LHC_mZphi}
\end{center}
\end{figure}

To better illustrate our argument from the previous paragraph, we show in Fig.~\ref{fig:LHC_mZphi} the $Z\phi$ invariant mass distribution for $\Lambda=4$~TeV, $\xi=1$, $m_{KK}=3 $~TeV, and two radion masses, $m_{\phi}=10$~GeV (left panel) and $m_{\phi}=750$~GeV (right panel). As the total cross section is obtained from the integration of the invariant mass distribution over values greater than the kinematical threshold, $\sqrt{\hat{s}}=m_{Z\phi} > m_{\phi}+m_Z$, it is clear why the KK $Z$ partners play a role only for the associated production of a heavy radion: in this case, the integral does not 
cover the region at low $\hat{s}$, where the invariant mass distribution is enhanced by the reduced ``off-shellness'' of the $Z$ boson contribution, thus giving more weight to the invariant mass region around the KK peak.

Moreover, one notices on the right panel of Fig.~\ref{fig:LHC_mZphi} that the two nearly-degenerate KK $Z$ bosons produce a single peak in the $Z\phi$ invariant mass distribution. In fact, as shown in this figure, this peak mostly originates from the $Z_2$ resonance, as it is, in general, more strongly coupled to $Z\phi$, than $Z_1$ is.
The other reason being that the $Z_1$ eigenstate 
is mainly composed of the $Z'$ boson which has vanishing couplings to the light initial quarks localised towards the Planck-brane. 
The interference term was taken at zero to draw those two resonance distributions separately. The spectacular observation of such a resonant $Z\phi$ production 
would represent the simultaneous direct manifestation of the radion and the first KK $Z$ boson, the rate of the extra boson $Z'$ (mainly constituting the $Z_1$ state) 
resonance being probably too small to expect a detection at LHC.

In addition, we have investigated the impact of varying the value of $g_R$ on the $Z\phi$ production cross section at the LHC. For this, we have chosen a point in the plane displayed in Fig.~\ref{fig:LHC_phiZ_mkk} and computed the corresponding  cross section for $g_R=g_L$ (Left-Right Parity case~\cite{O3})
and $g_R=2 g_L$ ($g_R\neq g_L$ is possible in different custodial symmetry implementations). Since one expects that changing $g_R$ would affect mostly the KK $Z$ bosons (not through small mixing effects, as is the case of $Z_0$), $Z_1$ and $Z_2$, we have considered $m_{\phi}=800$~GeV, such that the heavy KK resonances have a sizeable contribution to the $Z\phi$ production. Furthermore, we have taken $m_{KK}=3$~TeV and the other parameters as specified above the plot in Fig.~\ref{fig:LHC_phiZ_mkk}. The $Z\phi$ production cross sections for the two values of $g_R$ are of the same order of magnitude: while for $g_R=g_L$ we find $\sim 0.5$~fb, for $g_R= 2 g_L$ the cross section value is $\sim 0.15$~fb. The difference comes mostly from the $Z_i Z_0\phi$ ($i=1,2$) couplings, which are approximately two times stronger in the first case compared to the second case. The impact of the $g_R$ variation on the cross section is independent of the $\xi$ and $\Lambda$ parameters.

\begin{figure}[t]
\begin{center}
\includegraphics[keepaspectratio=true,width=0.46\textwidth]{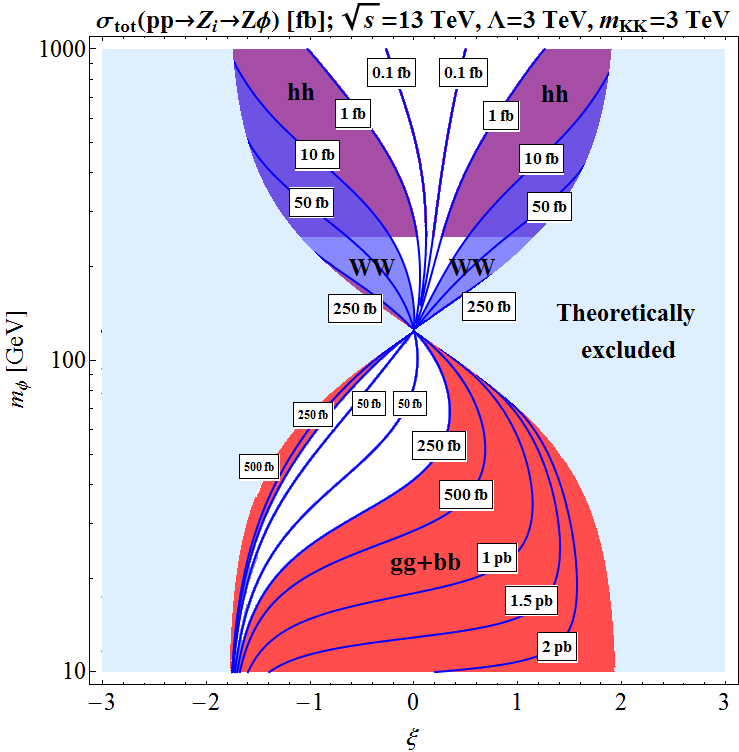}
\quad
\includegraphics[keepaspectratio=true,width=0.505\textwidth]{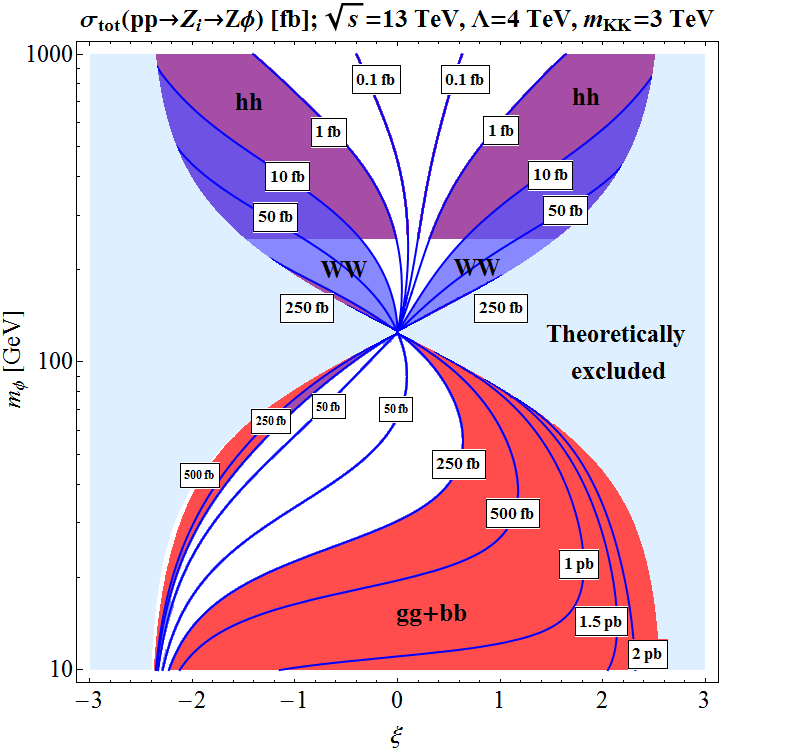}
\caption{
{\small Iso-contours of $Z\phi$ production cross section (in fb and pb) at the LHC with $\sqrt{s} = 13$~TeV, as a function of $\xi$ and $m_{\phi}$ (in GeV), for (left) $\Lambda=3$~TeV and (right) $\Lambda=4$~TeV
with $m_{KK}=3$~TeV. The light blue regions are excluded by the theoretical constraint from eq.~\eqref{th_excl}, while the purple, red, and blue zones approximately indicate parameter space regions that will be probed with $300$~fb$^{-1}$ at the LHC via radion decays into $hh$, dijets ($gg+bb$), and $WW$ final states, respectively.}
}
\label{fig:LHC_sigma_total}
\end{center}
\end{figure}

In Fig.~\ref{fig:LHC_sigma_total}, we present the total $Z\phi$ production cross section as a function of $\xi$ and $m_{\phi}$, for two values of the radion VEV, $\Lambda=3,4$~TeV, with $m_{KK}$ fixed at 3~TeV in both cases. We observe that for $m_{\phi}>m_h$ the cross section contours have roughly the same behaviours 
as the $g_{\phi}$ ones (see Fig.~\ref{fig:g_phi}). Indeed the dimensionless $g_{\phi}$ coupling
corresponds in a good approximation to the radion coupling to two Z bosons as described in the comments of Fig.~\ref{fig:g_phi}.
This is no longer true for $m_{\phi} < m_h$: in this latter region, as explained in the previous paragraphs, the cross section typically
increases as $m_{\phi}$ decreases, this being a result of the behaviour of PDFs, which increase at lower values of $\tau = \hat{s}/s$. However, even for $m_{\phi}<m_h$, the lowest $Z\phi$ production cross sections are achieved in the vicinity of the $g_{\phi}=0$ contour.

\subsubsection{Higgs Production}

\begin{figure}[t]
\begin{center}
\includegraphics[keepaspectratio=true,width=0.5\textwidth]{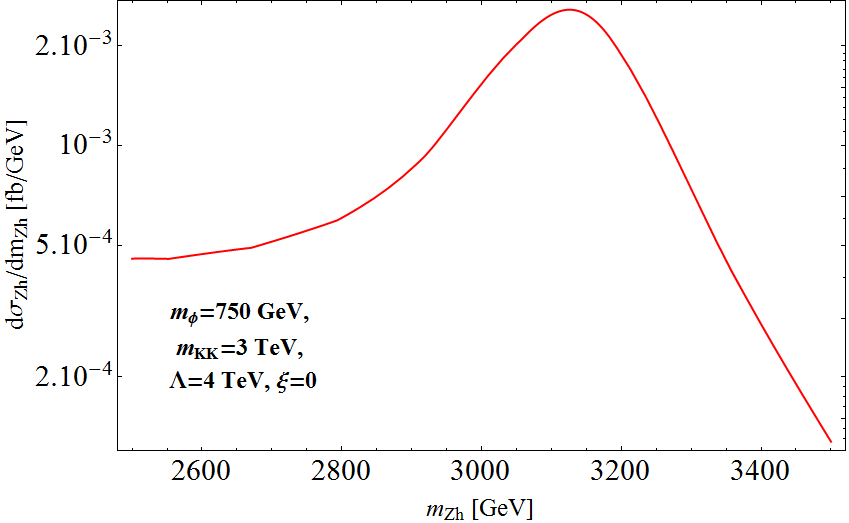}
\caption{
{\small $Zh$ invariant mass distribution (in fb/GeV) in the neighbourhood of the KK Z resonance peak, for $m_{\phi} = 750$~GeV. 
The values of the other relevant parameters are: $m_{KK} = 3$~TeV, $\Lambda = 4$~TeV, and $\xi=0$.}  
}
\label{fig:LHC_zh_peak}
\end{center}
\end{figure}

In Fig.~\ref{fig:LHC_zh_peak}, we show the $Zh$ invariant mass distribution, focusing on the region close to the resonant peak produced by the almost degenerate $Z_1$ and $Z_2$ states (the peak, as in the case of $Z\phi$ production, originates mostly from $Z_2$). We have chosen the following realistic parameters: 
$m_{\phi}=750$~GeV, $\Lambda=4$~TeV, $\xi=0$ and a mass of $m_{KK}=3$~TeV. 
The $Zh$ channel is a favoured discovery avenue for $Z_2$, as the largest branching ratio of $Z_2$ is into $Zh$ (meanwhile, $Z_1$ has its highest branching ratio for the $WW$ decay).
The observability potential for the KK resonance is discussed in Section~\ref{sec:KKresonance}.

\subsection{At the ILC}

We now focus our attention on the $Z\phi$ production at a linear electron-positron collider, taking as an example the International Linear Collider (ILC). For an $e^+ e^-$ collider, the problem is simpler, as the center-of-mass energy is a known quantity and one does not need to convolute the cross section with PDFs.

Another simplifying aspect is the fact that, for ILC center-of-mass energies, which in principle could go up to 1~TeV, the s-channel exchange of the KK partners of the $Z$ boson is negligible. Indeed as EWPT require that $m_{KK}$ is larger than $\sim 2 - 3$~TeV, the two heavy resonances, $Z_1$ and $Z_2$, are significantly off-shell even at $\sqrt{s}=1$~TeV, which renders their contribution negligible. Therefore, effectively, only the $Z$ boson exchange in the s-channel has to be considered for the $Z\phi$ production, as we have numerically checked. Concerning the KK $Z$ mixing effect on the $\phi ZZ$ coupling, for a given $Z \phi$ production cross section, varying $m_{KK}$ translates to at most a percent-level shifting of $\xi$ for a fixed $m_{\phi}$.

\begin{figure}[t]
\begin{center}
\includegraphics[keepaspectratio=true,width=0.48\textwidth]{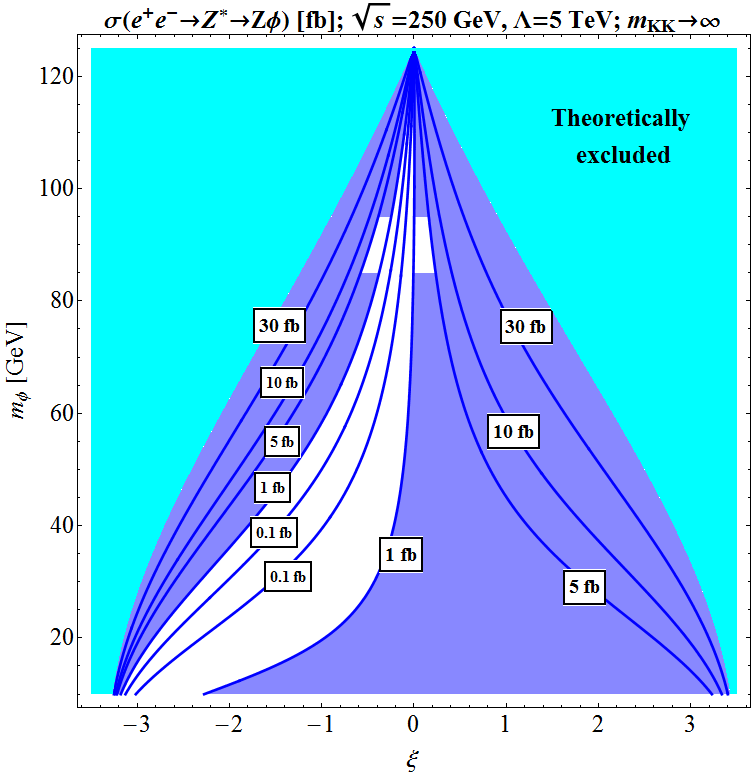}
\quad
\includegraphics[keepaspectratio=true,width=0.48\textwidth]{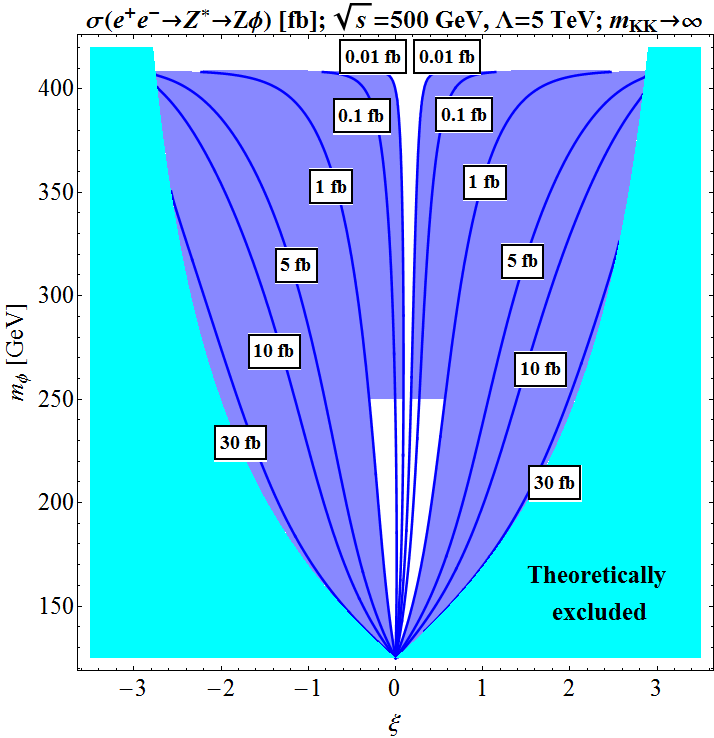}
\caption{
{\small Iso-contours of the $Z\phi$ production cross section (in fb) at the ILC with (left) $\sqrt{s} = 250$~GeV or (right) $\sqrt{s} = 500$~GeV, in terms of $\xi$ and $m_{\phi}$ (in GeV), for $\Lambda=5$~TeV and $m_{KK}\to\infty$. The cyan regions are excluded by the theoretical constraint from eq.~\eqref{th_excl}, while the blue zones indicate the 
parameter space regions estimated to be probed at the ILC through the $Z$ boson recoil mass technique.}
}
\label{fig:sigma_ilc}
\end{center}
\end{figure}

We plot in Fig.~\ref{fig:sigma_ilc} the $Z\phi$ production cross section in fb at the ILC, for $e^+ e^-$ center-of-mass energies of 250 and 500 GeV. We have chosen $\Lambda=5$~TeV and, to ease the calculations, $m_{KK}\to\infty$ (see previous paragraph). As described in Section~\ref{sec:AtLHCRad}, the hard process for the $Z\phi$ production cross section, as purely involved at the ILC (no PDF  effects), has typically the same dependence on the two parameters $\xi$ and $m_{\phi}$, as the $g_{\phi}$ coupling itself, whose values are illustrated on Fig.~\ref{fig:g_phi} (as a matter of fact, to a very good approximation, the aforementioned cross section is proportional to $g_{\phi}^2$). This explains the relative similarity of iso-contour behaviours between Fig.~\ref{fig:sigma_ilc} and Fig.~\ref{fig:g_phi} (upper left).

Notice that similarly to the SM $Zh$ production, the $Z\phi$ cross section, for a given radion mass, is proportional to $1/s$.~\footnote{Deviations from this behaviour are proportional to $g_{\phi}^r$, and in turn subdominant for most of the parameter space.} Consequently, in order to present the regions with maximal rates, we show in Fig.~\ref{fig:sigma_ilc} only small radion masses, $m_{\phi}<m_h$ for $\sqrt{s}=250$~GeV, while, for $\sqrt{s}=0.5$~TeV, we show only moderate to high radion masses, $m_{\phi}>m_h$.

\section{Radion, Higgs and KK Mode Detection} \label{sec:EXP}

\subsection{At the LHC} \label{sec:LHCapplied}

\subsubsection{Radion Decay to $b \bar b$} \label{sec:Rtobb}

For the full reaction $pp \to Z\phi$ followed by the radion decay into a
bottom quark pair, $\phi \to b \bar b$ (possibly including the decay
channel into two gluons),
the SM background comes from double gluon radiation in the process $q \bar
q \to Z$+2jets which has been well studied at LHC~\cite{DY2jets}.
At a 13 TeV LHC energy, the full rate for the Z boson production followed
by a muonic decay is $\sigma(pp\to Z)B(Z\to\mu^+\mu^-)\simeq 1900$~pb.

A drastic reduction of this background is therefore needed: it can come
from a cut on the transverse momentum of the reconstructed $Z$,
$p_T(\mu\mu)>100$~GeV
(see the $p_T(\mu\mu)$ distribution in Ref.~\cite{DY2jets}).
Such a cut would also induce a penalty on the $Z\phi$ production rate
approximately equivalent to imposing a cut on the $Z\phi$ invariant mass distribution,
$m_{Z\phi}>200$~GeV, which would lead to a drastic reduction factor of $1/40$ for example for the distribution
of Fig.~\ref{fig:LHC_mZphi} (left plot), obtained for a radion mass $m_{\phi} = 10$~GeV.
For heavier radions, $m_{\phi} \gtrsim 100$~GeV, the effect of this
optimal cut, $p_T(\mu\mu)>100$~GeV, is not significant since the $Z\phi$
invariant mass distribution
is defined on the range, $m_{Z\phi}>m_Z+m_{\phi}$.
A softer cut, $p_T(\mu\mu)>30$~GeV, would not alter significantly the
signal, even for $m_{\phi} = 10$~GeV,
and the background would be affected by a still efficient rejection factor
of $\sim 20$.

Let us now present guidelines on the main techniques to detect the $Z\phi$
production, depending on the radion mass.

$ \bullet $ \ $m_{\phi} \gtrsim 20$~GeV.
When $m_{\phi} \gtrsim 20$~GeV, it is justified to request two jets which
further decreases by an order of
magnitude the background (see Ref.~\cite{ATLAS2jets} for an ATLAS analysis
and Ref.~\cite{CMS2jets} for a CMS one).
Then a mass selection should gain a similar factor which brings us to a
rate of $\sim 1000$~fb for the background.
A bottom quark selection should gain an additional factor of
$10-100$~\cite{btag}.
Therefore, assuming a future integrated luminosity of $300$~fb$^{-1}$ at
the LHC, with a $20\%$ reconstruction efficiency on the signal and
background,
gives a $250$~fb sensitivity limit at $2\sigma$ on the cross section
$\sigma_{tot}(Z\phi)$, for a branching fraction $B(\phi \to  b \bar
b)\simeq 1$.
This corresponds to selecting experimentally two inclusive jets (including
two gluons or two $b$'s).
This LHC potential reach is illustrated on Fig.~\ref{fig:LHC_sigma_total}. 
On the obtained domains of the parameter space to which the LHC is potentially sensitive, one has indeed $B(\phi \to  b \bar b)\simeq 1$,
assuming standard
radion branching ratios without unknown physics entering the
radion-gluon-gluon triangular loop.
With $b$-tagging, the background should improve by about a factor
$2$ to $10$ (corresponding to a factor up to $\sqrt{10}$ in the limit),
depending on the tagging purity and efficiency, due to the further
background reduction.

$ \bullet $ \ $m_{\phi} > 100$~GeV.
At higher masses, say $m_{\phi} > 100$~GeV, the $p_T(\mu\mu)$ selection
cut can be increased up to $100$~GeV without damaging the signal
acceptance.
Besides, for these masses, the mass resolution increases and therefore the
sensitivity limit on $\sigma_{tot}(Z\phi)$ should reach about $100$~fb. 
This LHC potential reach covers higher mass regions in
Fig.~\ref{fig:LHC_sigma_total}.

\subsubsection{Radion Decay to $W^+W^-$} \label{sec:RtoWW}

$ \bullet $ \ $m_{\phi} > 160$~GeV.
In the regime $m_{\phi} > 160$~GeV, one benefits from the kinematical
opening of the $WW$ channel:
$pp \to Z\phi$, $\phi \to W^+W^-$~\footnote{One could as well benefit from
a cut on the transverse momentum of the reconstructed $Z$ based on such a
$p_T(\mu\mu)$ distribution for the associated $WWZ$ background.}.
The radion branching ratio into $ZZ$ is smaller.
The associated SM background composed of the $WWZ$ production has a cross
section of $\sim 200$~fb at 14 TeV including NLO QCD
corrections~\cite{ZWWrate}.
Assuming an integrated luminosity of $300$~fb$^{-1}$ at the LHC and
selecting semi-leptonic decays for the $WW$ system for a reconstruction
efficiency of $20\%$
(not including leptonic branching ratios), one expects $170$ events for
this SM background.
The radion mass selection then selects 20 events
corresponding to a $\sim 20$~fb sensitivity limit on the
$\sigma_{tot}(Z\phi)$ cross section, for a relevant branching $B(\phi \to 
W^+W^-)\simeq 0.5$; the associated sensitive region, for $m_{\phi} > 160$~GeV. This sensitivity order 
of magnitude is indicated on Fig.~\ref{fig:LHC_sigma_total}.

\subsubsection{Radion Decay to $hh$} \label{sec:Rtohh}

$ \bullet $ \ $m_{\phi} > 250$~GeV.
Finally, for $m_{\phi} > 250$~GeV, the LHC can become sensitive to the
channel $pp \to Z\phi$, $\phi \to hh$.
The $Zhh$ production background opens up with a cross section of
$0.25$~fb~\cite{Zhhrate}. Assuming a $20\%$ reconstruction efficiency,
including $b$-tagging,
would give a $0.5$ event background.
So $3$ events from the $Z\phi$ signal would be sufficient for a $2\sigma$
detection.
Hence one obtains a $\sim 5$~fb cross section sensitivity limit for
$\sigma_{tot}(Z\phi)$, with a realistic branching $B(\phi \to hh)\simeq
0.3$;
the corresponding domain, for $m_{\phi} > 250$~GeV. The order of magnitude of this sensitivity is indicated on
Fig.~\ref{fig:LHC_sigma_total} as well.

This domain and the above sensitivity regions are clearly coarse estimates
and a full analysis would be needed. Those regions however show that the
$Z\phi$ search
at LHC could be complementary, in testing some specific regions of the
$\{\xi,m_{\phi}\}$ plane, to the search for the gluon-gluon fusion
mechanism of radion
production, in case this loop-induced process is not affected by an
unknown physics underlying the SM: this mechanism allows to cover large
domains of the RS
parameter space as shown in the figures of Ref.~\cite{Frank} (regions
below $m_{\phi} = 80$~GeV were not studied there).

\subsubsection{KK Resonances} \label{sec:KKresonance}

The $Z\phi$ production can exhibit degenerate KK mode resonances made of $Z$
boson excitations as described in Section~\ref{sec:AtLHC}.
These resonances show up in the bump of Fig.~\ref{fig:LHC_mZphi}. In order to discuss the possibility of a KK resonance observation in the radion production,
we now consider some optimised but realistic parameter values, $\Lambda=3$~TeV,
$\xi=1.5$, and $m_{\phi}= 500$~GeV (see the upper left plot of Fig.~\ref{fig:g_phi}).
Then the integrated rate of such a resonant process, obtained by considering an interval $m_{Z_2}\pm 2 \, \Gamma_{Z_2}$ on Fig.~\ref{fig:LHC_mZphi}, 
is of $\sim 10$~fb ($\sim 1$~fb)  for $m_{KK}= 2 $~TeV (3~TeV). For a (HL-)LHC luminosity of $300(0)$~fb$^{-1}$, the induced
number of events might lead to a possible but challenging observation.
The kinematic selection of the interval around $m_{KK}$ in the $Z\phi$
invariant mass distribution would reduce the associated SM background.
The $p_T(\mu\mu)$ selection cut keeps a good efficiency if the production
of $Z\phi$ is dominated by the exchange of a KK $Z$ resonance.
For $m_{KK}\simeq 2$~TeV and a radion mass below $\sim 120$~GeV, a simple
kinematical study shows that a cut $p_T(\mu\mu)\gtrsim 1$~TeV
would select the signal peaked in this area while eliminating
significantly the QCD background.
A complete Monte Carlo simulation of the signal and background would be
needed to conclude on the observability of such a resonance.

This $p_T(\mu\mu)$ selection method is generic and can even be applied for
the various processes of the type $q \bar q \to Y\to XZ$ where $Y$ is a
heavy vector
boson which can be produced on-shell and $X$ is a lighter resonance, either
SM-like ($W, Z, h$) or exotic, as is the case for the radion. An additional
advantage of this
process is that it provides a combination of two resonances allowing a
double discrimination. In this respect, the LHC could be competitive with
ILC where the
production of an on-shell $Y$ resonance is only possible for a mass
$m_Y<1$~TeV.

Similarly, the $Zh$ production can occur through KK $Z$ boson resonances as
shown in Fig.~\ref{fig:LHC_zh_peak}. For the optimised parameter values, $\Lambda=4$~TeV, $\xi=1$, $m_{\phi}=500$~GeV (see the upper right plot of Fig.~\ref{fig:g_phi}), and an optimistic low mass $m_{KK}\simeq
2$~TeV,
the obtained integrated rate is of $\sim 11.5$ fb. Similar remarks as for the
$Z\phi$ production hold regarding the KK resonance observability.

\subsubsection{Higgs Production} \label{sec:HiggsProdLHC}

The Higgs coupling to two $Z$ bosons has been measured at the LHC, via the
Higgs production in association with a $Z$ boson.
Assuming decoupling KK modes (which do not affect significantly the
$Z\phi$ production), the Higgs couplings are modified only by the
Higgs-radion mixing.
Taking this into account, the experimental values for the $hZZ$ coupling
exclude some domains of the $\{\xi,m_{\phi}\}$ plane. However, as we shall see later on in Section~\ref{sec:HiggsProdILC}, these domains are not significant when compared to the ILC sensitivity.  

\begin{figure}[t]
\begin{center}
\includegraphics[keepaspectratio=true,width=0.48\textwidth]{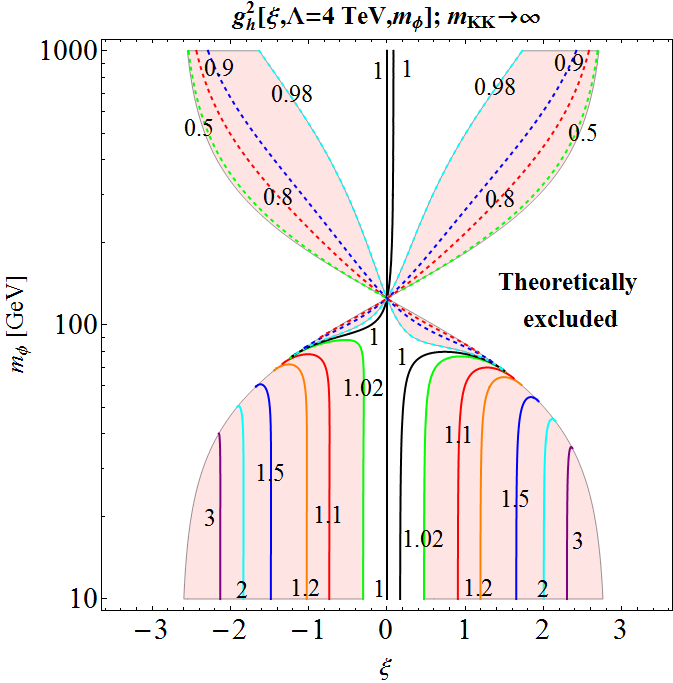}
\quad
\includegraphics[keepaspectratio=true,width=0.48\textwidth]{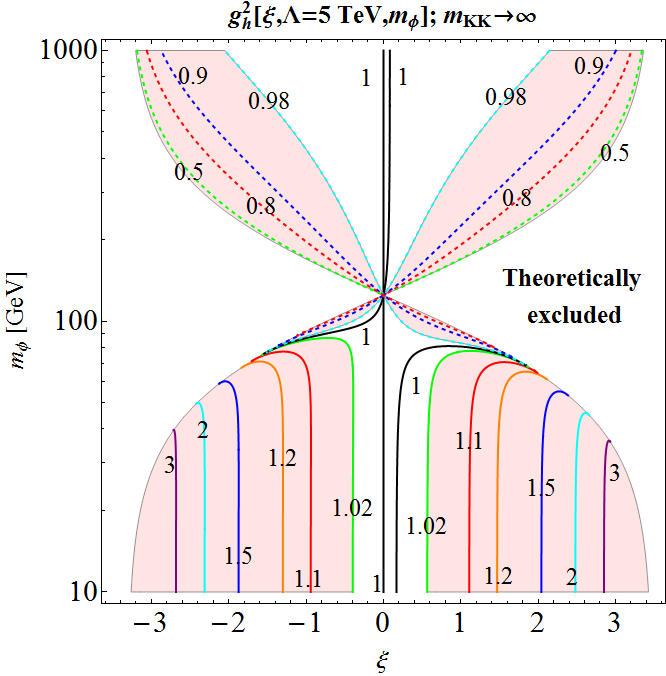}
\caption{
{\small Iso-contours of $g_h^2$ in the $\{\xi,m_{\phi}\}$ plane, for (left)
$\Lambda=4$~TeV and (right) $\Lambda=5$~TeV, with $m_{KK}$ taken to
infinity. The coloured region indicates the future indirect sensitivity of the ILC on the Higgs-radion parameter space, 
corresponding to a $\sim 2\%$ accuracy (at $2\sigma$) on the measurement of the squared $hZZ$ coupling, i.e. $0.98<g_h^2<1.02$.}
}
\label{fig:g_h_sq}
\end{center}
\end{figure}

A first LHC analysis combines the run 1 measurements (ATLAS and CMS)~\cite{atlas_cms_higgs}, with global fits reporting a central value of $\sim 1$ (i.e. SM value) and a $\sim 10\%$ error at 1$\sigma$ on $g_h$ (defined in eq.~\eqref{g_phi}~\footnote{In the $m_{KK}\to\infty$ limit employed here (where $C_0^{\rm 4D}\to 1$), $g_h$ represents indeed the $hZZ$ coupling normalised to its SM value, since the second term in eq.~\eqref{phi_ZZ} is vanishing in this limit and the third one is more than $2$ orders of magnitude smaller.} and denoted by $\kappa_Z$ in Ref.~\cite{atlas_cms_higgs}), assuming that the Higgs decays only into SM states. Therefore, in our case, this constraint is relevant only for $m_{\phi} >  m_h/2$. Moreover, it allows for $0.6<g_h^2<1.4$ at $2\sigma$, which covers a tiny region in the $g_h^2$ plot from Fig.~\ref{fig:g_h_sq}.

Ref.~\cite{atlas_cms_higgs} also presents global fits allowing for Higgs boson decays to non-SM states, but with the extra assumption that $g_h <1$ (or $\kappa_Z<1$ in their notation), which is not justified in our framework. Their result indicates that, at two sigma, $0.6<g_h^2<1$, which means that, once again, only a tiny region from Fig.~\ref{fig:g_h_sq} is covered. 

Even though, regarding the $hZZ$ coupling measurement, the LHC is much less competitive than the ILC, these exclusions can still be seen as a new interpretation of the constraints on the
RS model from the LHC Higgs data, in the presence of a Higgs-radion mixing
(see also Ref.~\cite{HRconst}). The Higgs physics appears naturally as
complementary to the radion sector in testing their common
$\{\xi,m_{\phi}\}$ parameter space.

\subsection{At the ILC} \label{sec:ILCapplied}

\subsubsection{Radion Production} \label{sec:RadionProdILC}

For the associated $Z\phi$ production at ILC, one can use the same missing
mass technique as for the $Zh$ production~\cite{MissMass} which is
independent of
the radion branching ratio values.
This powerful method is only feasible using the large luminosity provided
by this machine which plans to collect
$2000$~fb$^{-1}$ at $250$~GeV (H-20 scenario~\cite{H20scen}),
$4000$~fb$^{-1}$ at $500$~GeV and $8000$~fb$^{-1}$ at $1$~TeV.
This is to be compared to the LEP collider which could only collect a few
fb$^{-1}$ per experiment so that LEP was
not able to significantly exclude the presence of a radion at any mass.
This recoil mass technique works best near the $Z\phi$ threshold where the
center-of-mass energy is about $m_\phi+m_Z$. One then achieves the most
precise
recoil mass reconstruction. For this reason the low mass domain, $m_{\phi}
\lesssim 160$~GeV, will be covered by running at a center-of-mass energy
of $250$~GeV.

$ \bullet $ \ $m_{\phi} < m_Z$.
When $m_{\phi} < m_Z$, one has an easy situation. The Z background from
$ZZ^*/\gamma^*$ is distributed as a Breit Wigner with a small tail at low
masses due to the
virtual photon contribution from $Z\gamma^*$.
The sensitivity reaches a limit on the $\sigma(Z\phi)$ of $\sim
1$~fb at the $2\sigma$ statistical level.
When the $b\bar b$ decay mode is considered, this sensitivity limit goes
even down to $0.02$~fb.

$ \bullet $ \ $m_{\phi} \sim m_Z$.
For $m_{\phi} \sim m_Z$, the ZZ background is the largest but still giving
a sensitivity limit on $\sigma(Z\phi)$ of $\sim 3$~fb at $2\sigma$.

$ \bullet $ \ $m_{\phi} > m_Z$.
If $m_{\phi} > m_Z$, one ends up with a similar situation as for $Zh$: the
main background comes from $ZZ$+ISR, where ISR stands for
initial state
radiation (i.e. a photon radiated off $e^+/e^-$) which, in most cases, remains undetected. The
missing mass however includes both the Z and this photon, creating what
one calls a radiative
tail (for $m_{\phi} \sim m_Z$, the mass reconstruction of the Z into
hadrons is too imprecise to allow a separation of $m_{\phi}$ from $m_Z$).
From Ref.~\cite{MissMass},  one can easily evaluate the
$\sigma(Z\phi)$ sensitivity in this mass region which is at the
$1$~fb level.
The $Zh$ channel itself creates a background which generates a small blind
zone for $m_{\phi} \simeq m_h$ but in this case the Higgs properties
can also be altered allowing one to feel the presence of the radion.

$ \bullet $ \ $m_{\phi} > 130$~GeV.
At $m_{\phi} >130$~GeV, it becomes possible to eliminate the radiative
tail effect by reconstructing the radion mass through its decays into two
jets. The
$\sigma(Z\phi)$ sensitivity improves to $0.5$~fb.

$ \bullet $ \ $m_{\phi} > 150$~GeV.
When $m_{\phi} >150$~GeV, one starts crossing the kinematical limit for
the $Z\phi$ production and it becomes necessary to use data taken at a
$500$~GeV
center-of-mass energy. The recoil mass precision is poor since one
operates far above the $Z\phi$ threshold, but the good energy resolution
on jets
($\sigma E_j/E_j\sim 3\%$) allows to use direct mass reconstruction with a
mass resolution on the radion at the $2\%$ level. One can then include the
leptonic and
neutrino decay modes from $Z$, gaining a factor $\sim 10$ in efficiency.
Since one is no more suffering from the ISR effect this method turns out
to give a sensitivity for
$\sigma(Z\phi)$ at the $0.1$~fb level.

$ \bullet $ \ $m_{\phi} > 160$~GeV.
For $m_{\phi} >160$~GeV, the situation changes radically since the $WW$, $ZZ$
channels become accessible for the radion decay, which helps the recoil
techniques.
For the SM background, the Ref.~\cite{Doublet} on $WWZ$ cross sections
shows that the $WWZ$ contribution can be reduced down to $10$~fb by using
right-handed polarization ($e_R$) for the electron beam. The SM $ZZZ$
background is at the $1$~fb level. For $ZWW$ one can simply use the
$Z\to\mu\mu$ tagging. The $WW$
component can be identified through semi-leptonic decays where a $W$ decays
hadronically and the other leptonically. Taking into account the branching
ratios,
one expects $350$ background events. At the counting level one reaches a
$1$~fb sensitivity on $\sigma(Z\phi)$. One can then select the
$\phi$ mass
allowing an increased sensitivity of about $0.3$~fb.

$ \bullet $ \ $m_{\phi} > 250$~GeV.
For $m_{\phi} >250$~GeV, the $hh$ channel becomes accessible for the radion
decay. The $Zhh$ SM background~\cite{TDRILC} is even smaller and with
strong signatures given by the Higgs decay into $b\bar b$. Assuming a
$50\%$ efficiency with a relevant $B(\phi\to hh)\sim 0.3$ and low extra
backgrounds (from $ZZZ$ essentially), one could reach a sensitivity on $\sigma(Z\phi)$
at the $0.01$~fb level.
For the other ILC option with a $1$~TeV center-of-mass energy and an
integrated luminosity of $8000$~fb$^{-1}$, the factor increase in
luminosity, compared to the $500$~GeV scenario, induces a factor
$\sqrt{2}$ of improvement in the cross section sensitivity (the $Zhh$
background is only slightly smaller).

\begin{figure}[t]
\begin{center}
\includegraphics[keepaspectratio=true,width=0.48\textwidth]{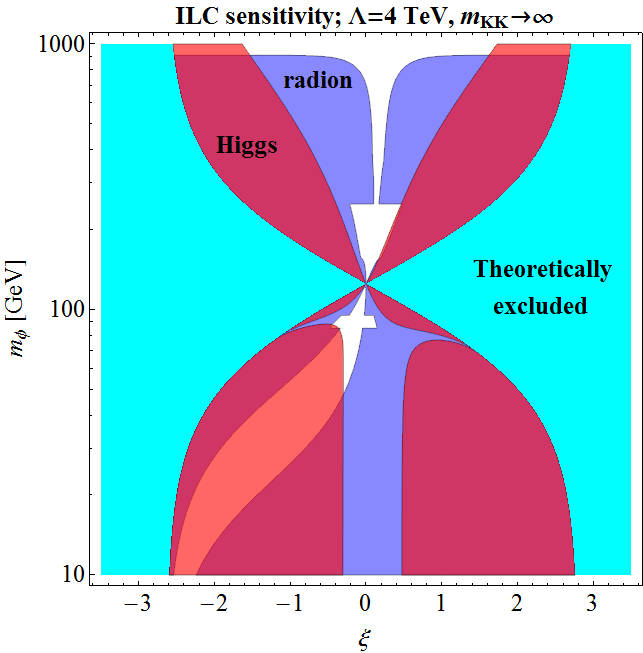}
\quad
\includegraphics[keepaspectratio=true,width=0.48\textwidth]{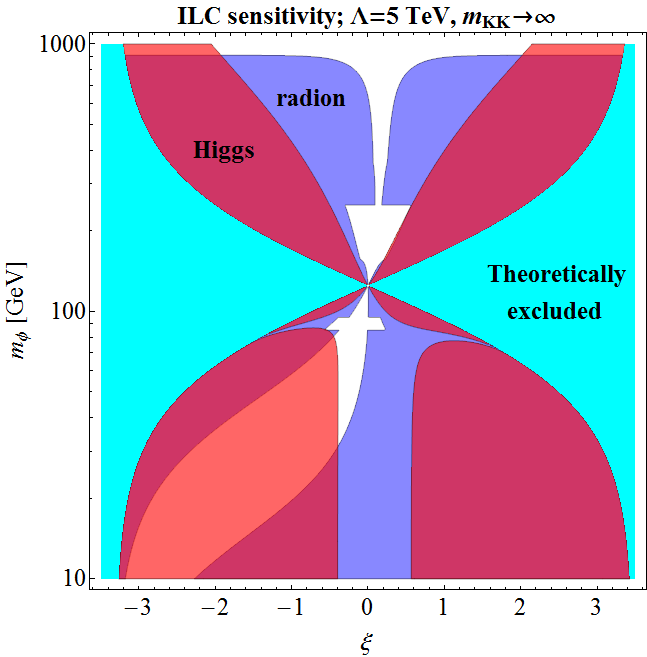}
\caption{
{\small Summary plots for direct and indirect radion searches at the three stages of operation of the ILC ($\sqrt{s} = 250$~GeV, $500$~GeV, and $1$~TeV), in the $\{\xi,m_{\phi}\}$ plane, for (left) $\Lambda=4$~TeV and (right) $\Lambda=5$~TeV, with $m_{KK}$ taken to be infinite. The blue region covers the Higgs-radion parameter space estimated to be probed by the ILC through direct radion searches, while the red region represents the domain potentially probed by the precise measurement of the $hZZ$ coupling. The theoretical constraint is superimposed once more, as the cyan
domain.}
}
\label{fig:summary_ilc}
\end{center}
\end{figure}

The various estimates given so far constitute a reasonable first guess of
the ILC sensitivity for a radion search. All the obtained orders of magnitude for the sensitivities on
$\sigma(Z\phi)$
given in the text are drawn as indicative coloured regions in
Fig.~\ref{fig:sigma_ilc}. On Fig.~\ref{fig:summary_ilc}, we summarize on a
unique plot
the covered regions issued from two possible ILC runs respectively at
$250$~GeV, $500$~GeV and $1$~TeV, for infinite $m_{KK}$ (i.e. decoupled KK resonances) and two values of the radion VEV, $\Lambda=4,5$~TeV.
A dedicated analysis would be needed to fully assess such performances but
it is clear that ILC can dig into the radion scenario with excellent
sensitivity.

We notice that the region corresponding to $\xi = 0$ and $m_{\phi}\simeq 60 - 110$~GeV, left uncovered on Fig.~\ref{fig:summary_ilc}, 
might be tested via the search for the reaction $gg\to\phi\to\gamma\gamma$ at the HL-LHC extension with an integrated 
luminosity of $3000$~fb$^{-1}$: this is the conclusion of Ref.~\cite{LightRadgg} in the case of SM fields localised on the TeV-brane.

Besides, as for the SM Higgs case, the vector boson fusion mechanism could provide additional information on the radion,
in particular allowing the determination of the total width and in turn of absolute widths~\cite{TDRILC}.

\subsubsection{Higgs Production} \label{sec:HiggsProdILC}

The Higgs coupling to two $Z$ bosons would possibly be measured at the
$0.51\%$ ($1.3\%$) $1\sigma$ error level at the ILC with an energy option
of $1$~TeV ($250$~GeV), for a luminosity of $2500$~fb$^{-1}$
($250$~fb$^{-1}$)~\cite{ILCperH},
via the Higgs production in association with a $Z$ boson.
Such measurements would exclude at $2\sigma$ the regions of the
$\{\xi,m_{\phi}\}$ plane, as illustrated in Fig.~\ref{fig:g_h_sq}, assuming a
central value equal
to the predicted SM $hZZ$ coupling constant. Notice that this measurement is
independent of the Higgs branching ratio values due to the recoil technique
used to tag the associated $Z$ boson. The future precision Higgs physics at ILC 
would thus be extremely efficient in testing the $\{\xi,m_{\phi}\}$ parameter space, 
as illustrated in Fig.~\ref{fig:g_h_sq}.
The obtained exclusion regions are superimposed as well on the summary
plot of Fig.~\ref{fig:summary_ilc} showing, the whole parameter space than
can
be covered using both the $Z\phi$ and $Zh$ production at ILC.

\section{Conclusion}  \label{se:conclu}

Let us finish this study on the radion production by a short conclusion, now that the numerical results have been discussed in detail
with respect to the possibilities of observation.
The investigation of the reaction $q \bar q \to Z\phi$ at LHC could allow to cover significant parts of the RS parameter space. 
This reaction could even benefit from the resonance of degenerate neutral KK vector bosons, which would enhance the reaction and allow for 
tight selections against the QCD background. It will take the ILC program at high luminosity to cover most of the theoretically allowed 
parameter space, via the $e^+e^- \to Z\phi$ search. The ILC, via such a reaction investigation, is particularly complementary of the LHC for testing the low radion masses 
(below the Higgs mass) since the reaction $gg\to\phi\to\gamma\gamma$ is quite efficient in principle to probe the high mass regime. The ILC benefits from the complementarity, 
of the direct radion searches and the high accuracy measurements of the Higgs couplings, in the exploration of the RS parameter space (typically the $\{\xi,m_{\phi}\}$ plane).

\vspace*{1cm}

\section*{Acknowledgements}

The authors gratefully acknowledge A.Ahmed, A.Djouadi, F.Nortier and B.Rossignol for stimulating discussions.
The work of A.A. is supported by the ERC advanced grant ``Higgs@LHC''.
G.M. would like to thank the support from the CNRS LIA (Laboratoire International Associ\'e), THEP (Theoretical High Energy
Physics) and the INFRE-HEPNET (IndoFrench Network on High Energy Physics) of
CEFIPRA/IFCPAR (Indo-French Centre for the Promotion of Advanced Research).

\end{document}